\documentclass[article,nojss]{jss}
\usepackage[utf8]{inputenc}
\graphicspath{{Figures/}}

\providecommand{\tightlist}{%
  \setlength{\itemsep}{0pt}\setlength{\parskip}{0pt}}

\author{
Achim Zeileis\\Universität Innsbruck \And Jason C. Fisher\\U.S. Geological Survey \And Kurt Hornik\\WU Wirtschafts-\\
universität Wien \And Ross Ihaka\\University of Auckland \AND Claire D. McWhite\\The University of\\
Texas at Austin \And Paul Murrell\\University of Auckland \And Reto Stauffer\\Universität Innsbruck \And Claus O. Wilke\\The University of\\
Texas at Austin
}
\title{\pkg{colorspace}: A Toolbox for Manipulating and Assessing Colors and
Palettes}

\Plainauthor{Zeileis, Fisher, Hornik, Ihaka, McWhite, Murrell, Stauffer, Wilke}
\Plaintitle{colorspace: A Toolbox for Manipulating and Assessing Colors and Palettes}
\Shorttitle{\pkg{colorspace}: Manipulating and Assessing Colors and Palettes}

\Abstract{
The \proglang{R} package \pkg{colorspace} provides a flexible toolbox
for selecting individual colors or color palettes, manipulating these
colors, and employing them in statistical graphics and data
visualizations. In particular, the package provides a broad range of
color palettes based on the HCL (Hue-Chroma-Luminance) color space. The
three HCL dimensions have been shown to match those of the human visual
system very well, thus facilitating intuitive selection of color
palettes through trajectories in this space.
Using the HCL color model general strategies for three types of palettes are implemented:
(1)~Qualitative for coding categorical information, i.e., where no
particular ordering of categories is available. (2)~Sequential for
coding ordered/numeric information, i.e., going from high to low (or
vice versa). (3)~Diverging for coding ordered/numeric information around
a central neutral value, i.e., where colors diverge from neutral to two
extremes.
To aid selection and application of these palettes the package also
contains scales for use with \pkg{ggplot2}, \pkg{shiny} (and
\pkg{tcltk}) apps for interactive exploration, visualizations of palette
properties, accompanying manipulation utilities (like desaturation and
lighten/darken), and emulation of color vision deficiencies.
}

\Plainkeywords{color, palette, HCL, RGB, hue, color vision deficiency, R}

\Address{
  Achim Zeileis\\
  Universität Innsbruck\\
  Department of Statistics\\
  Faculty of Economics and Statistics\\
  Universitätsstr. 15\\
  6020 Innsbruck, Austria\\
  E-mail: \email{Achim.Zeileis@R-project.org}\\
  URL: \url{https://eeecon.uibk.ac.at/~zeileis/}\\~\\
}

\usepackage{thumbpdf,lmodern}
\usepackage{booktabs,amsmath,longtable,supertabular,framed}
\shortcites{color:colorspace,color:ggplot2pkg}

\begin{document}

\vspace*{0.6cm}

\section{Introduction}\label{sec:intro}

Color is an integral and omnipresent element of many statistical
graphics and data visualizations. Therefore, colors should be carefully
chosen to support all viewers in accessing the information displayed
\citep{color:Tufte:1990, color:Brewer:1999, color:Ware:2004, color:Wilkinson:2005, color:Wilke:2019}.
However, until relatively recently many software packages have been
using color palettes derived from simple RGB (red-green-blue) color
combinations such as the RGB ``rainbow'' (or ``jet'') color palette with
poor perceptual properties. See
\citet{color:Hawkins+McNeall+Stephenson:2014} and
\citet{color:Stauffer+Mayr+Dabernig:2015} and the references therein for
an overview.

To address these problems, many improved color palettes with better
perceptual properties have been receiving increasing attention in the
literature
\citep{color:Harrower+Brewer:2003, color:Zeileis+Hornik+Murrell:2009, color:Smith+VanDerWalt:2015, color:CARTO, color:Crameri:2018}.
Many systems for systems for statistical and scientific computing
provide infrastructure for such color palettes, e.g., for \proglang{R}
\citep{color:R} the list of useful packages encompasses:
\pkg{RColorBrewer} \citep{color:RColorBrewer}, \pkg{viridis}
\citep{color:viridis}, \pkg{rcartocolor} \citep{color:rcartocolor},
\pkg{wesanderson} \citep{color:wesanderson}, or \pkg{scico}
\citep{color:scico} among many others. Furthermore, packages like
\pkg{pals} \citep{color:pals} and \pkg{paletteer}
\citep{color:paletteer} collect many of the proposed palettes in
combination with a unified interface. Most of these palettes, however,
are pre-existing palettes, stored as a limited set of colors and
interpolated as necessary. And even if specific algorithms have been
used in the initial construction of the palettes, these are often not
reflected in the software implementations.

\begin{figure}[t!]
\centering
\includegraphics[width=0.5\textwidth]{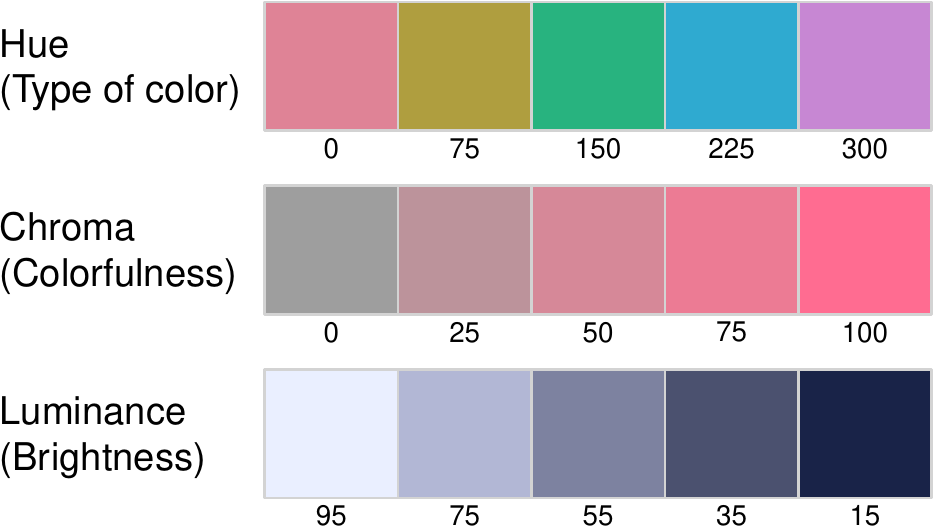} 
\caption[Axes of the HCL color space]{Axes of the HCL color space. Top: Hue $H$ changes from 0 (red) via 75 (yellow), etc.\ to 300 (purple) with fixed $C = 60$ and $L = 65$. Center: Chroma $C$ changes from 0 (gray) to 100 (colorful) with fixed $H = 0$ (red) and $L = 65$. Bottom: Luminance $L$ changes from 95 (light) to 15 (dark) with fixed $H = 260$ (blue) and $C = 25$ (low, close to gray).}\label{fig:hcl-properties}
\end{figure}

The \pkg{colorspace} package \citep{color:colorspace} adopts a somewhat
different approach that gives the user direct access to the construction
principles underlying its palettes. These are based on simple
trajectories in the perceptually-based HCL (Hue--Chroma--Luminance)
color space \citep{color:Wiki+HCL} whose axes match those of the human
visual system very well: Hue (= type of color, dominant wavelength),
chroma (= colorfulness), luminance (= brightness), see
Figure~\ref{fig:hcl-properties}. Thus, utilizing this color model the
\pkg{colorspace} package can derive general and adaptable strategies for
color palettes; manipulate individual colors and color palettes; and
assess and visualize the properties of color palettes (beyond simple
color swatches). Specifically, \pkg{colorspace} provides three types of
palettes based on the HCL model:
\begin{itemize}
\tightlist
\item
  \emph{Qualitative:} Designed for coding categorical information, i.e.,
  where no particular ordering of categories is available and every
  color should receive the same perceptual weight. Function:
  \texttt{qualitative\_hcl()}.
\item
  \emph{Sequential:} Designed for coding ordered/numeric information,
  i.e., where colors go from high to low (or vice versa). Function:
  \texttt{sequential\_hcl()}.
\item
  \emph{Diverging:} Designed for coding ordered/numeric information
  around a central neutral value, i.e., where colors diverge from
  neutral to two extremes. Function: \texttt{diverging\_hcl()}.
\end{itemize}
A broad collection of prespecified palettes are shipped in the package.
In addition, existing palettes can be easily tweaked and new or adapted
palettes registered. The prespecified palettes include suitable HCL
color choices that closely approximate of most palettes from packages
\pkg{RColorBrewer}, \pkg{rcartocolor}, and \pkg{viridis} by using only a
small set of hue, chroma, and luminance parameters.

To aid choice and application of these palettes the package provides (a)
scales for use with \pkg{ggplot2} \citep{color:ggplot2}, (b) \pkg{shiny}
\citep{color:shiny} and \pkg{tcltk} \citep{color:R} apps for interactive
exploration, (c) visualizations of palette properties, and (d)
accompanying manipulation utilities (like desaturation, lighten/darken,
and emulation of color vision deficiencies).

The remainder of the manuscript is organized as follows:
Section~\ref{sec:tour} gives a first overview of the package's ``look \&
feel'' and the general workflow. Section~\ref{sec:color_spaces}
summarizes the \proglang{S}4 color space classes and methods in the
package. Section~\ref{sec:hcl_palettes} introduces the extensible
collection of HCL-based palettes along with their construction details.
Section~\ref{sec:palette_visualization} presents the toolbox for palette
visualization and assessment. Section~\ref{sec:color_vision_deficiency}
discusses the implemented techniques for color vision deficiency
emulation that help assess the suitability of colors for colorblind
viewers. Section~\ref{sec:hclwizard} briefly highlights the interactive
color apps from the package. Finally, some further color manipulation
utilities are highlighted in Section~\ref{sec:manipulation_utilities}
before Section~\ref{sec:summary} concludes the manuscript.

\section{A quick tour}\label{sec:tour}

The stable release version of \pkg{colorspace} is hosted on the
Comprehensive \proglang{R} Archive Network (CRAN) at
\url{https://CRAN.R-project.org/package=colorspace} and the development
version of \pkg{colorspace} is hosted on \proglang{R}-Forge at
\url{https://R-Forge.R-project.org/projects/colorspace/}.

\subsection{Choosing HCL-based color palettes}\label{choosing-hcl-based-color-palettes}

\begin{figure}[t!]
\centering
\includegraphics[width=\textwidth]{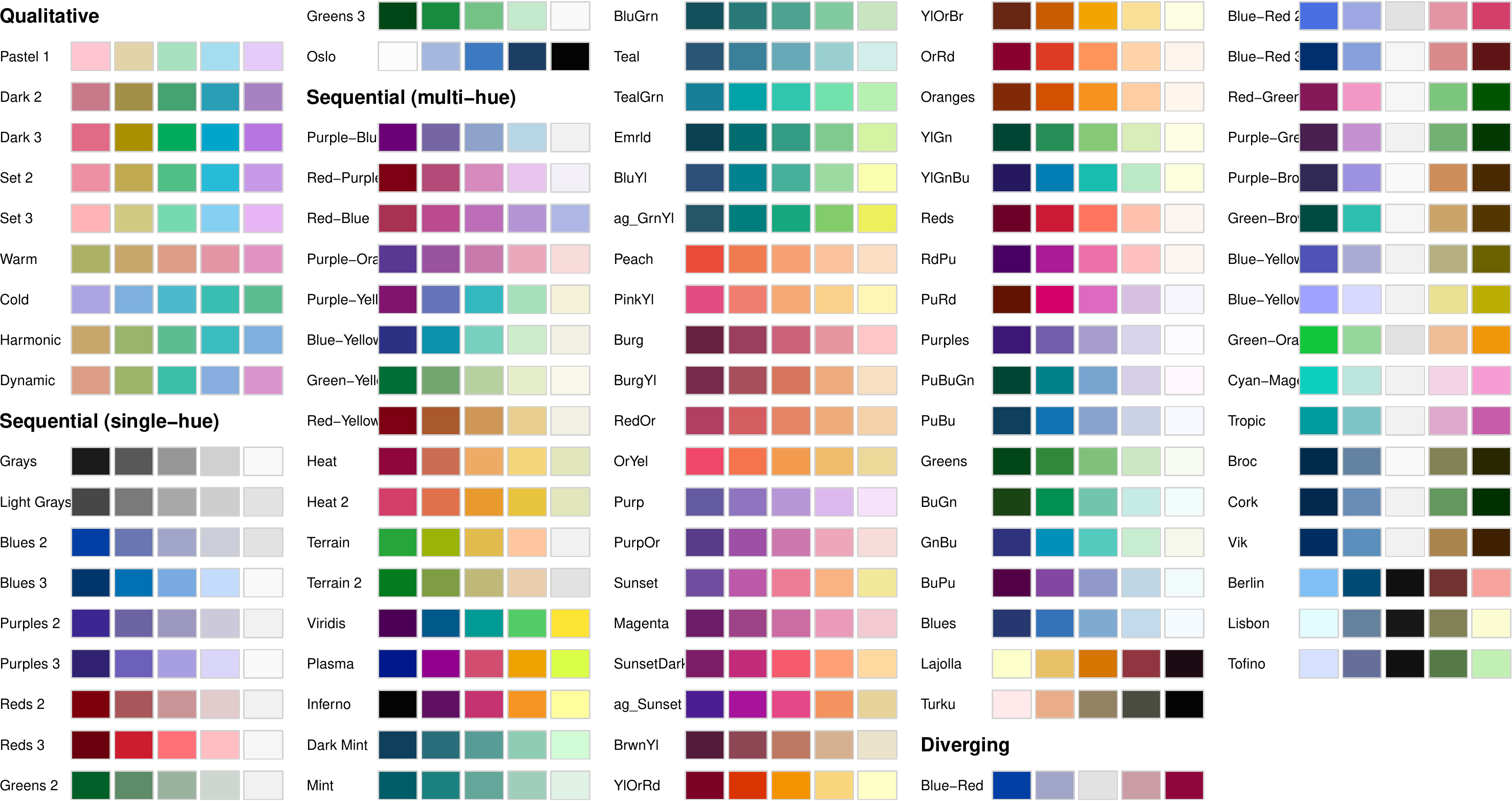} 
\caption[Brief overview of available predefined palettes in \pkg{colorspace}]{Brief overview of available predefined palettes in \pkg{colorspace}.}\label{fig:hcl-palettes}
\end{figure}

The \pkg{colorspace} package ships with a wide range of predefined color
palettes, specified through suitable trajectories in the HCL
(Hue-Chroma-Luminance) color space. A quick overview can be gained
easily with the \texttt{hcl\_palettes()} function (see
Figure~\ref{fig:hcl-palettes}):
\begin{CodeChunk}
\begin{CodeInput}
R> library("colorspace")
R> hcl_palettes(plot = TRUE)
\end{CodeInput}
\end{CodeChunk}
A suitable vector of colors can be easily computed by specifying the
desired number of colors and the palette name (see
Figure~\ref{fig:hcl-palettes} for possible palette names), e.g.,
\begin{CodeChunk}
\begin{CodeInput}
R> q4 <- qualitative_hcl(4, palette = "Dark 3")
R> q4
\end{CodeInput}
\begin{CodeOutput}
[1] "#E16A86" "#909800" "#00AD9A" "#9183E6"
\end{CodeOutput}
\end{CodeChunk}
The functions \texttt{sequential\_hcl()}, and \texttt{diverging\_hcl()}
work analogously. Additionally, a palette's hue/chroma/luminance
parameters can be modified, thus allowing for easy customization of each
palette. Moreover, the \texttt{choose\_palette()}/\texttt{hclwizard()}
app provides convenient user interfaces to perform palette customization
interactively. Finally, even more flexible diverging HCL palettes are
provided by \texttt{divergingx\_hcl()}.

\subsection{Usage with base graphics}\label{usage-with-base-graphics}

\begin{figure}[t!]
\centering
\includegraphics[width=\textwidth]{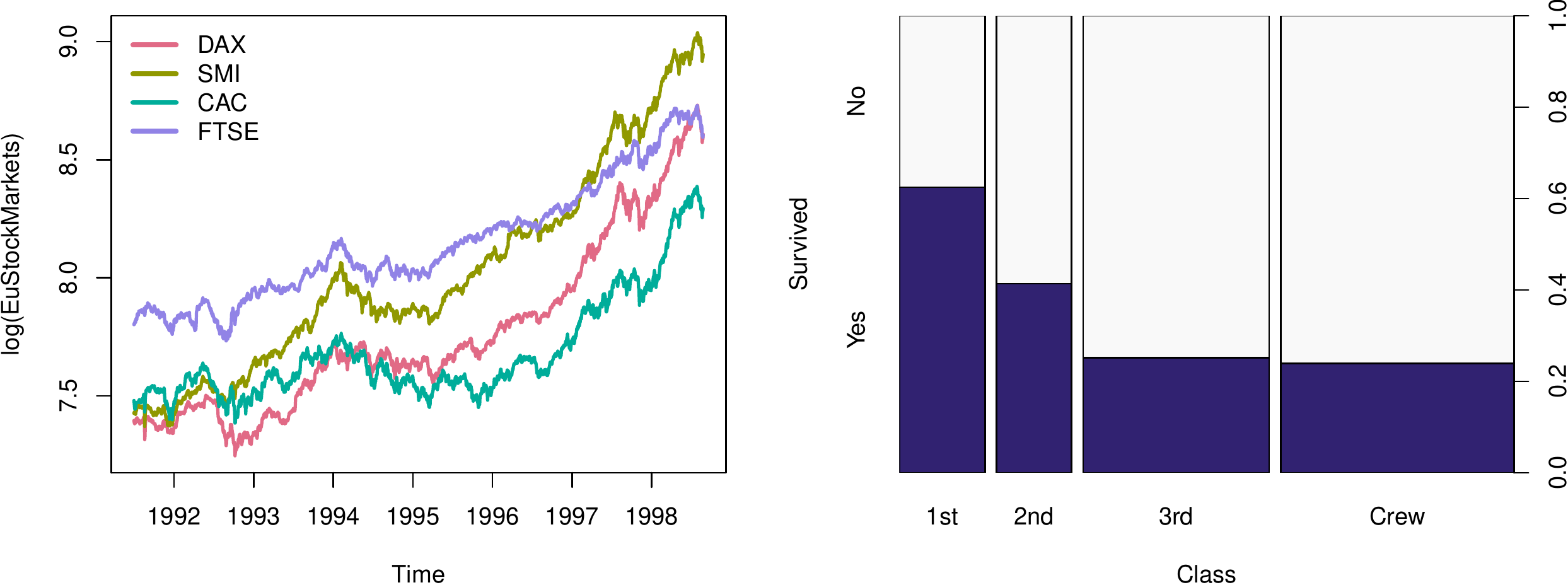} 
\caption{Using \pkg{colorspace} with base \proglang{R} graphics. Left: Time series plot of log-prices from \code{EuStockMarkets} data with \code{qualitative\_hcl(4, "Dark 3")} palette. Right: Spine plot with survival proportions across passenger classes in the \code{titanice} data with \code{sequential\_hcl(2, "Purples 3")} palette.}\label{fig:eustockmarkets-titanic}
\end{figure}

The color vectors returned by the HCL palette functions can usually be
passed directly to most base graphics function, typically through the
\texttt{col} argument. Here, the \texttt{q4} vector created above is
used in a time series display (see the left panel of
Figure~\ref{fig:eustockmarkets-titanic}):
\begin{CodeChunk}
\begin{CodeInput}
R> plot(log(EuStockMarkets), plot.type = "single", col = q4, lwd = 2)
R> legend("topleft", colnames(EuStockMarkets), col = q4, lwd = 3, bty = "n")
\end{CodeInput}
\end{CodeChunk}
As another example for a sequential palette, we demonstrate how to
create a spine plot (see the right panel of
Figure~\ref{fig:eustockmarkets-titanic}) displaying the proportion of
Titanic passengers that survived per class. The \texttt{"Purples\ 3"}
palette is used, which is quite similar to the \pkg{ColorBrewer.org}
\citep{color:Harrower+Brewer:2003} palette \texttt{"Purples"}. Here,
only two colors are employed, yielding a dark purple and light gray.
\begin{CodeChunk}
\begin{CodeInput}
R> ttnc <- margin.table(Titanic, c(1, 4))[, 2:1]
R> spineplot(ttnc, col = sequential_hcl(2, palette = "Purples 3"))
\end{CodeInput}
\end{CodeChunk}

\subsection[Usage with ggplot2]{Usage with \pkg{ggplot2}}\label{sec:ggplot2}

\begin{figure}[t!]
\centering
\includegraphics[width=0.49\textwidth]{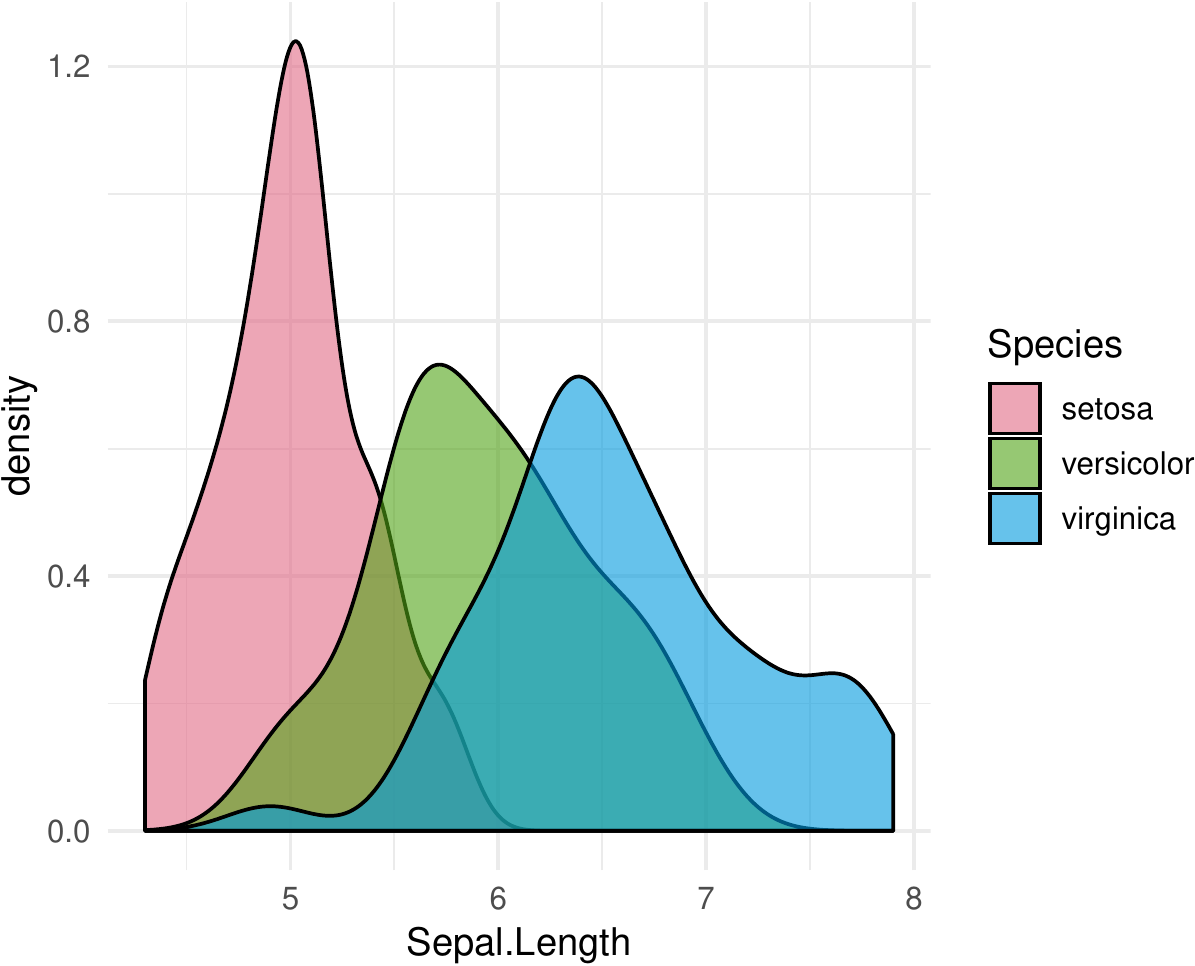} \includegraphics[width=0.49\textwidth]{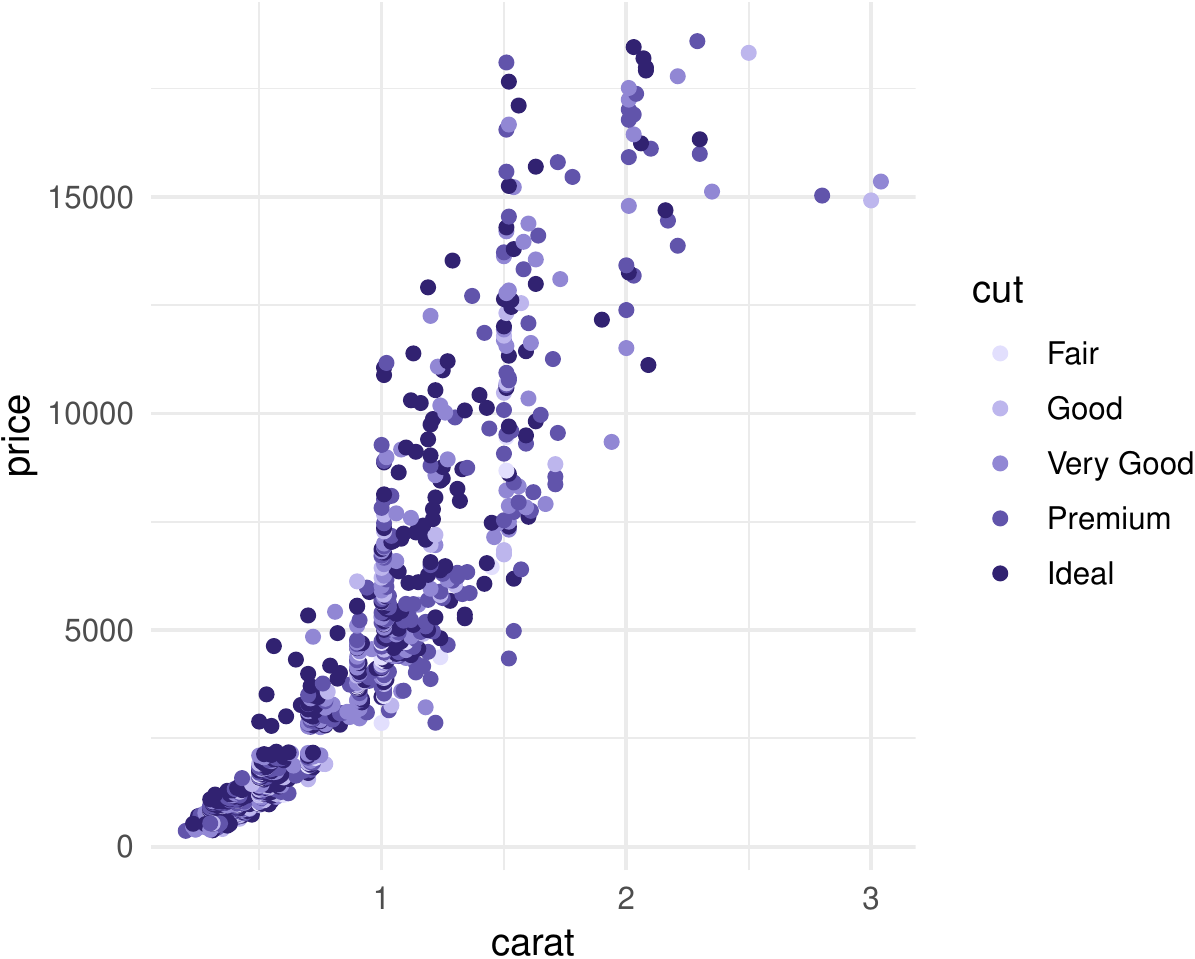} 
\caption{Using \pkg{colorspace} with \pkg{ggplot2} graphics. Left: Kernel density of sepal length, grouped and shaded by species, in the \code{iris} data with semi-transparent \code{scale\_fill\_discrete\_qualitative(palette = "Dark 3")} color scale. Right: Scatter plot of price by carat, shaded by cut levels, in a subsample of the \code{diamonds} data with the \code{scale\_color\_discrete\_sequential(palette = "Purples 3", nmax = 6, order = 2:6)} color scale.}\label{fig:iris-diamonds}
\end{figure}

To provide access to the HCL color palettes from within \pkg{ggplot2}
graphics \citep{color:ggplot2, color:ggplot2pkg} suitable discrete
and/or continuous \pkg{gglot2} color scales are provided. The scales are
named via the scheme
\begin{verbatim}
scale_<aesthetic>_<datatype>_<colorscale>()
\end{verbatim}
where
\begin{itemize}
\tightlist
\item
  \texttt{\textless{}aesthetic\textgreater{}} is the name of the
  aesthetic (\texttt{fill}, \texttt{color}, \texttt{colour}).
\item
  \texttt{\textless{}datatype\textgreater{}} is the type of the variable
  plotted (\texttt{discrete} or \texttt{continuous}).
\item
  \texttt{\textless{}colorscale\textgreater{}} sets the type of the
  color scale used (\texttt{qualitative}, \texttt{sequential},
  \texttt{diverging}, \texttt{divergingx}).
\end{itemize}
To illustrate their usage two simple examples are shown using the
qualitative \texttt{"Dark\ 3"} and sequential \texttt{"Purples\ 3"}
palettes that were also employed above. For the first example,
semi-transparent shaded densities of the sepal length from the
\texttt{iris} data are shown, grouped by species (see the left panel of
Figure~\ref{fig:iris-diamonds}).
\begin{CodeChunk}
\begin{CodeInput}
R> library("ggplot2")
R> ggplot(iris, aes(x = Sepal.Length, fill = Species)) +
+    geom_density(alpha = 0.6) +
+    scale_fill_discrete_qualitative(palette = "Dark 3")
\end{CodeInput}
\end{CodeChunk}
And for the second example the sequential palette is used to code the
cut levels in a scatter of price by carat in the \texttt{diamonds} data
(or rather a small subsample thereof, see the right panel of
Figure~\ref{fig:iris-diamonds}). The scale function first generates six
colors but then drops the first color because the light gray is too
light here. (Alternatively, the chroma and luminance parameters could
also be tweaked.)
\begin{CodeChunk}
\begin{CodeInput}
R> dsamp <- diamonds[1 + 1:1000 * 50, ]
R> ggplot(dsamp, aes(carat, price, color = cut)) + geom_point() +
R>   scale_color_discrete_sequential(palette = "Purples 3",
R>     nmax = 6, order = 2:6)
\end{CodeInput}
\end{CodeChunk}

\subsection{Palette visualization and assessment}\label{palette-visualization-and-assessment}

\begin{figure}[t!]
\centering
\includegraphics[width=\textwidth]{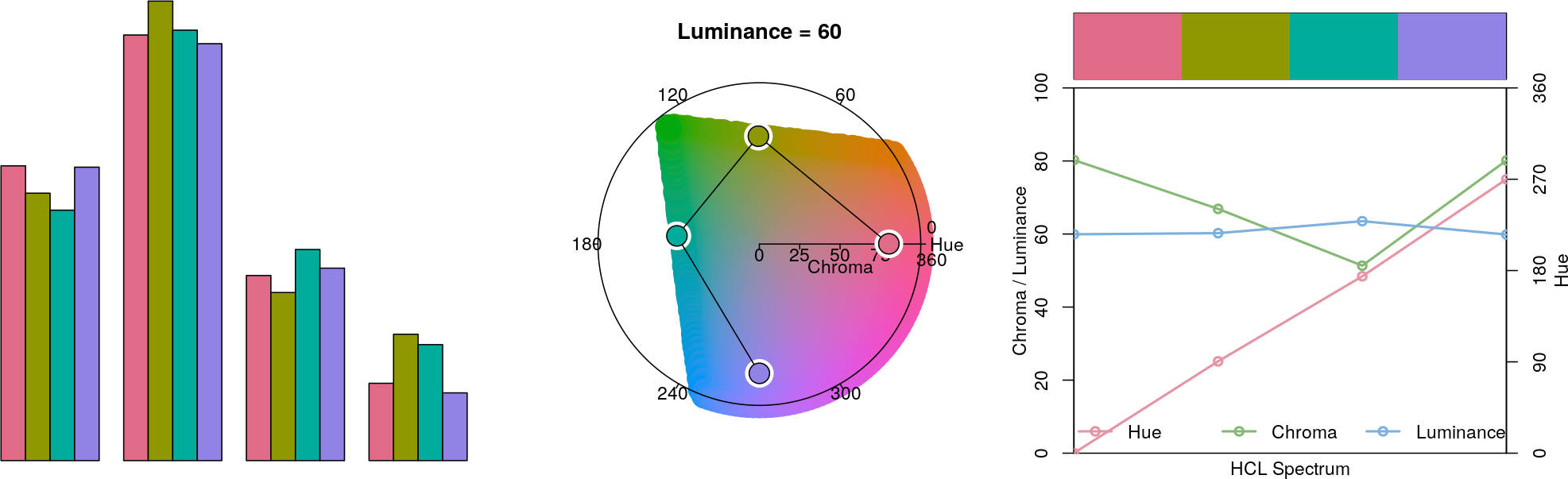} 
\caption[Palette visualization and assessment for \code{qualitative\_hcl(4, "Dark 3")} palette]{Palette visualization and assessment for \code{qualitative\_hcl(4, "Dark 3")} palette. Left: Demo bar plot. Center: Hue-chroma plane at fixed $L = 60$ in HCL space. Right: HCL spectrum with linearly changing hue (around color wheel), almost constant chroma, and constant luminance.}\label{fig:allplots-qualitative}
\end{figure}

The \pkg{colorspace} package also provides a number of functions that
aid visualization and assessment of its palettes.
\begin{itemize}
\tightlist
\item
  \texttt{demoplot()} can display a palette (with arbitrary number of
  colors) in a range of typical and somewhat simplified statistical
  graphics.
\item
  \texttt{hclplot()} converts the colors of a palette to the
  corresponding hue/chroma/luminance coordinates and displays them in
  HCL space with one dimension collapsed. The collapsed dimension is the
  luminance for qualitative palettes and the hue for
  sequential/diverging palettes.
\item
  \texttt{specplot()} also converts the colors to hue/chroma/luminance
  coordinates but draws the resulting spectrum in a line plot.
\end{itemize}
For the qualitative \texttt{"Dark\ 3"} palette from above the following
plots can be obtained (see Figure~\ref{fig:allplots-qualitative}).
\begin{CodeChunk}
\begin{CodeInput}
R> demoplot(q4, "bar")
R> hclplot(q4)
R> specplot(q4, type = "o")
\end{CodeInput}
\end{CodeChunk}
The bar plot is used as a typical application for a qualitative palette
(in addition to the time series and density plots used above). The other
two displays show that luminance is (almost) constant in the palette
while the hue changes linearly along the color ``wheel''. Ideally,
chroma would have also been constant to completely balance the colors.
However, at this luminance the maximum chroma differs across hues so
that the palette is fixed up to use less chroma for the yellow and green
elements.

Note also that in a bar plot areas are shaded (and not just points or
lines) so that lighter colors would be preferable. In the density plot
in Figure~\ref{fig:iris-diamonds} this was achieved through
semi-transparency. Alternatively, luminance could be increased as is
done in the \texttt{"Pastel\ 1"} or \texttt{"Set\ 3"} palettes.

\begin{figure}[t!]
\centering
\includegraphics[width=\textwidth]{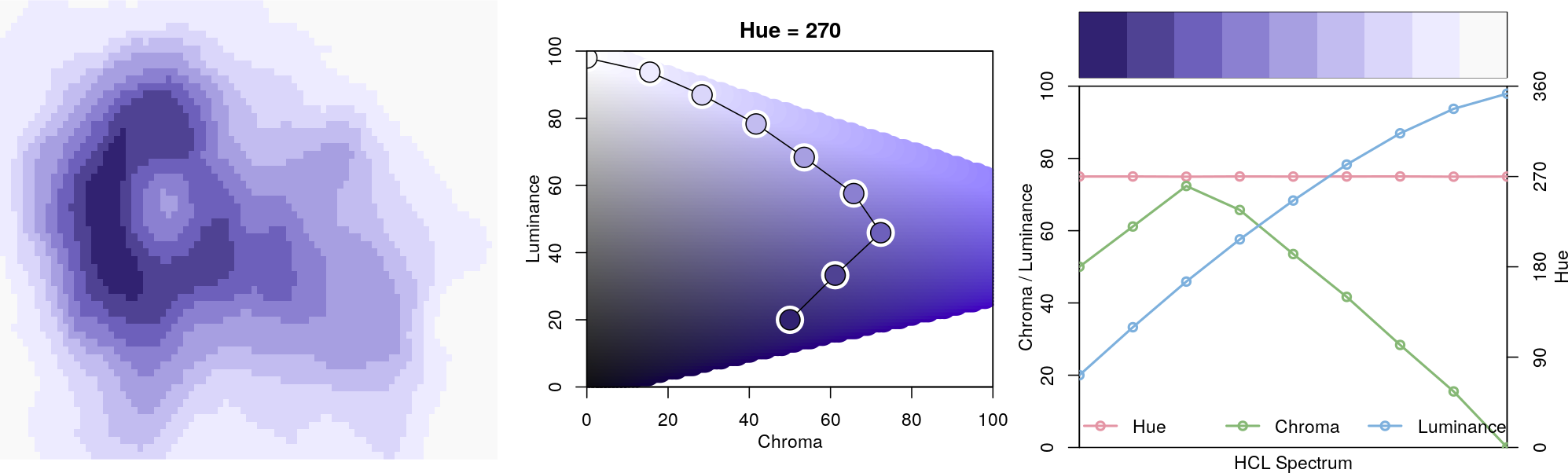} 
\caption[Palette visualization and assessment for \code{sequential\_hcl(4, "Purples 3")} palette]{Palette visualization and assessment for \code{sequential\_hcl(4, "Purples 3")} palette. Left: Demo heatmap. Center: Chroma-luminance plane at fixed $H = 270$ in HCL space. Right: HCL spectrum with constant hue, triangular chroma, and increasting luminance.}\label{fig:allplots-sequential}
\end{figure}

Subsequently, the same types of assessment are carried out in
Figure~\ref{fig:allplots-sequential} for the sequential
\texttt{"Purples\ 3"} palette as employed above.
\begin{CodeChunk}
\begin{CodeInput}
R> s9 <- sequential_hcl(9, "Purples 3")
R> demoplot(s9, "heatmap")
R> hclplot(s9)
R> specplot(s9, type = "o")
\end{CodeInput}
\end{CodeChunk}
In Figure~\ref{fig:allplots-sequential}, a heatmap (based on the
well-known Maunga Whau volcano data) is used as a typical application
for a sequential palette. The elevation of the volcano is brought out
clearly, using dark colors to give emphasis to higher elevations. The
other two displays show that hue is constant in the palette while
luminance and chroma vary. Luminance increases monotonically from dark
to light (as required for a proper sequential palette). Chroma is
triangular-shaped which allows to better distinguish the middle colors
in the palette when compared to a monotonic chroma trajectory.

\pagebreak

\section[Color spaces: S4 classes and utilities]{Color spaces: \proglang{S}4 classes and utilities}\label{sec:color_spaces}

\begin{figure}[b!]
\centering
\includegraphics[width=0.9\textwidth]{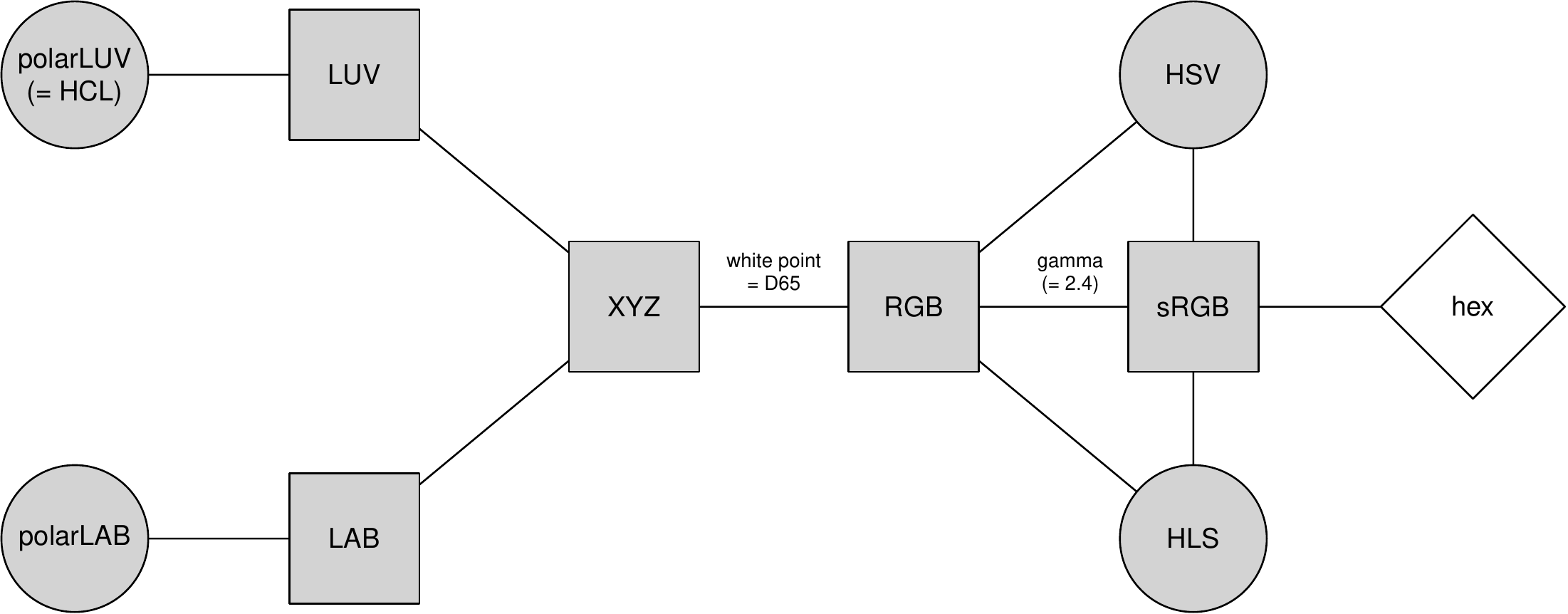} 
\caption[Relationship between three-dimensional color spaces implemented in \pkg{colorspace}]{Relationship between three-dimensional color spaces implemented in \pkg{colorspace}. Color models that are (or try to be) perceptually-based are displayed with circles, other color models with rectangles.}\label{fig:color-spaces}
\end{figure}

At the core of the \pkg{colorspace} package are various utilities for
computing with color spaces \citep{color:Wiki+Colorspace}, as the name
conveys. Thus, the package helps to map various three-dimensional
representations of color to each other \citep{color:Ihaka:2003}. A
particularly important mapping is the one from the perceptually-based
and device-independent color model HCL (Hue-Chroma-Luminance) to
standard Red-Green-Blue (sRGB) which is the basis for color
specifications in many systems based on the corresponding hex codes
\citep{color:Wiki+Webcolors}, e.g., in HTML but also in \proglang{R}.
For completeness further standard color models are included as well in
the package. The connections are illustrated in
Figure~\ref{fig:color-spaces}. Color models that are (or try to be)
perceptually-based are displayed with circles and models that are not
are displayed with rectangles.

\subsection{Implemented color spaces}\label{implemented-color-spaces}

The color spaces, implemented in \pkg{colorspace}, along with their
corresponding \proglang{S}4 classes and eponymous class constructors,
are:
\begin{itemize}
\tightlist
\item
  \texttt{RGB()} for the classic Red-Green-Blue color model that mixes
  three primary colors with different intensities to obtain a spectrum
  of colors. The advantage of this color model is (or was) that it
  corresponded to how computer and TV screens generated colors, hence it
  was widely adopted and still is the basis for color specifications in
  many systems. For example, the hex color codes are employed in HTML
  but also in \proglang{R}. However, the RGB model also has some
  important drawbacks: It does not take into account the output device
  properties, it is not perceptually uniform (a unit step within RGB
  does not produce a constant perceptual change in color), and it is
  unintuitive for humans to specify colors (say brown or pink) in this
  space. See \citet{color:Wiki+RGB} for more details.
\item
  \texttt{sRGB()} addresses the issue of device dependency by adopting a
  so-called gamma correction. Therefore, the gamma-corrected standard
  RGB (sRGB), as opposed to the linearized RGB above, is a good model
  for specifying colors in software and for hardware. But it is still
  unintuitive for humans to work directly with this color space.
  Therefore, sRGB is a good place to end up in a color space
  manipulation but it is not a good place to start. See
  \citet{color:Wiki+sRGB} for more details.
\item
  \texttt{HSV()} is a simple transformation of the (s)RGB space that
  tries to capture the perceptual axes: \emph{hue} (dominant wavelength,
  the type of color), \emph{saturation} (colorfulness), and \emph{value}
  (brightness, i.e., light vs.~dark). Unfortunately, the three axes in
  the HSV model are confounded so that, e.g., brightness changes
  dramaticaly with hue. See \citet{color:Wiki+HSV} for more details.
\item
  \texttt{HLS()} (Hue-Lightness-Saturation) is another transformation of
  (s)RGB that tries to capture the perceptual axes. It does a somewhat
  better job but the dimensions are still strongly confounded. See
  \citet{color:Wiki+HSV} for more details.
\item
  \texttt{XYZ()} was established by the CIE (Commission Internationale
  de l'Eclairage) based on psychophysical experiments with human
  subjects. It provides a unique triplet of XYZ values, coding the
  standard observer's perception of the color. It is device-independent
  but it is not perceptually uniform and the XYZ coordinates have no
  intuitive meaning. See \citet{color:Wiki+CIEXYZ} for more details.
\item
  \texttt{LUV()} and \texttt{LAB()} were therefore proposed by the CIE
  as perceptually uniform color spaces where the former is typically
  preferred for emissive technologies (such as screens and monitors)
  whereas the latter is usually preferred when working with dyes and
  pigments. However, the three axes of these two spaces still do not
  correspond to human perceptual axes. See \citet{color:Wiki+CIELUV};
  \citet{color:Wiki+CIELAB} for more details.
\item
  \texttt{polarLUV()} and \texttt{polarLAB()} take polar coordinates in
  the UV plane and AB plane, respectively. Specifically, the polar
  coordinates of the LUV model are known as the HCL
  (Hue-Chroma-Luminance) model. These capture the human perceptual axes
  very well without confounding effects as in the HSV or HLS approaches.
  (More details follow below.)
\end{itemize}
All \proglang{S}4 classes for color spaces inherit from a virtual class
\texttt{color} which is internally always represented by matrices with
three columns (corresponding to the different three dimensions).

Note that since the inception of the color space conversion tools within
\pkg{colorspace}'s \proglang{C} code by \citet{color:Ihaka:2003} other
\proglang{R} tools for this purpose became available. Notably, base
\proglang{R} meanwhile provides \texttt{grDevices::convertColor()}
\citep[computed in high-level \proglang{R},][]{color:R} and
\texttt{farver::convert\_colour()} \citep{color:farver} is based on a
\proglang{C++} library. For many basic color conversion purposes the
\pkg{colorspace} package and these alternatives are essentially equally
suitable \citep[see the discussion in][]{color:convertColor}. For more
complex conversions, including different chromatic adaptation
algorithms, a more comprehensive color science approach is implemented
in the \proglang{R} package \pkg{colorscience}
\citep{color:colorscience}. Finally, base \proglang{R} also provides
\texttt{grDevices::hcl()} for mapping HCL representations to hex codes.

To make the \pkg{colorspace} package self-contained and exactly backward
compatible, the \proglang{C} code in \pkg{colorspace} is still used as
the basis for all color space conversions.

\subsection{Utilities}\label{utilities}

For working with the implemented \proglang{S}4 color spaces various
utilities are available:
\begin{itemize}
\tightlist
\item
  \texttt{as()} method: Conversions of a \texttt{color} object to the
  various color spaces, e.g., \texttt{as(x,\ "sRGB")}.
\item
  \texttt{coords()}: Extract the three-dimensional coordinates
  pertaining to the current color class.
\item
  \texttt{hex()}: Convert a color object to sRGB and code in a hex
  string that can be used within \proglang{R} plotting functions.
\item
  \texttt{hex2RGB()}: Convert a given hex color string to an sRGB color
  object which can also be coerced to other color spaces.
\item
  \texttt{readRGB()} and \texttt{readhex()} can read text files into
  color objects, either from RGB coordinates or hex color strings.
\item
  \texttt{writehex()}: Writes hex color strings to a text file.
\item
  \texttt{whitepoint()}: Query and change the white point employed in
  conversions from CIE XYZ to RGB. Defaults to D65.
\end{itemize}

\subsection[Illustration of basic colorspace functionality]{Illustration of basic \pkg{colorspace} functionality}\label{illustration-of-basic-colorspace-functionality}

As an example a vector of colors \texttt{x} can be specified in the HCL
(or polar LUV) model:
\begin{CodeChunk}
\begin{CodeInput}
R> (x <- polarLUV(L = 70, C = 50, H = c(0, 120, 240)))
\end{CodeInput}
\begin{CodeOutput}
      L  C   H
[1,] 70 50   0
[2,] 70 50 120
[3,] 70 50 240
\end{CodeOutput}
\end{CodeChunk}
The resulting three colors are pastel red (hue = 0), green (hue = 120),
and blue (hue = 240) with moderate chroma and luminance. For display in
other systems an sRGB representation might be needed:
\begin{CodeChunk}
\begin{CodeInput}
R> (y <- as(x, "sRGB"))
\end{CodeInput}
\begin{CodeOutput}
             R         G         B
[1,] 0.8931564 0.5853740 0.6465459
[2,] 0.5266113 0.7224335 0.4590469
[3,] 0.4907804 0.6911937 0.8673877
\end{CodeOutput}
\end{CodeChunk}
With \texttt{coords(x)} or \texttt{coords(y)} the displayed coordinates
can also be extracted as numeric matrices. And from sRGB we can also
coerce to HSV for example:
\begin{CodeChunk}
\begin{CodeInput}
R> as(y, "HSV")
\end{CodeInput}
\begin{CodeOutput}
            H         S         V
[1,] 348.0750 0.3446008 0.8931564
[2,] 104.6087 0.3645825 0.7224335
[3,] 208.0707 0.4341857 0.8673877
\end{CodeOutput}
\end{CodeChunk}
For display in many systems (including \proglang{R} itself) hex color
codes based on the sRGB coordinates can be created:
\begin{CodeChunk}
\begin{CodeInput}
R> hex(x)
\end{CodeInput}
\begin{CodeOutput}
[1] "#E495A5" "#86B875" "#7DB0DD"
\end{CodeOutput}
\end{CodeChunk}

\section{HCL-based color palettes}\label{sec:hcl_palettes}

\begin{figure}[t!]
\centering
\includegraphics[width=\textwidth]{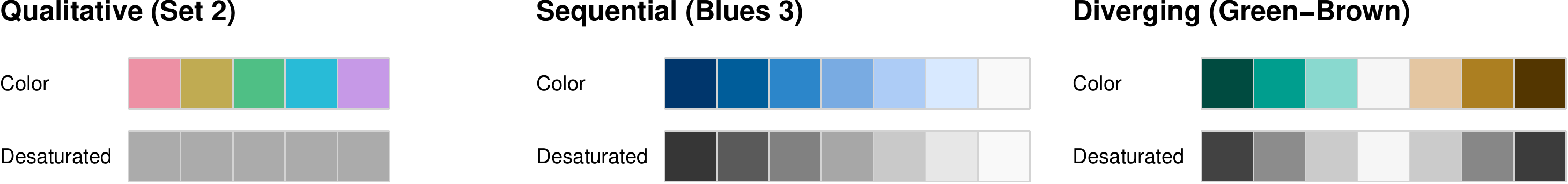} 
\caption[Examples of palette types in \pkg{colorspace}]{Examples of palette types in \pkg{colorspace}. Qualitative palettes are balanced towards the same luminance level while sequential and diverging palettes go from dark to light and/or vice versa, respectively.}\label{fig:hcl-palettes-principles}
\end{figure}

As motivated in the previous section, the HCL space is particularly
useful for specifying individual colors and color palettes, as its three
axes match those of the human visual system very well. Therefore, the
\pkg{colorspace} package provides three palette functions based on the
HCL model: \texttt{qualitative\_hcl()}, \texttt{sequential\_hcl()}, and
\texttt{diverging\_hcl()}. Their construction principles are exemplified
in Figure~\ref{fig:hcl-palettes-principles} and explained in more detail
below. The desaturated palettes in the second row of
Figure~\ref{fig:hcl-palettes-principles} bring out clearly that
luminance differences (= light-dark contrasts) are crucial for
sequential and diverging palettes while qualitative palettes are
balanced at the same luminance.

To facilitate obtaining good sets of colors, HCL parameter combinations
that yield useful palettes are accessible by name. These can be listed
using the function \texttt{hcl\_palettes()}:
\begin{CodeChunk}
\begin{CodeInput}
R> hcl_palettes()
\end{CodeInput}
\begin{CodeOutput}
HCL palettes

Type:  Qualitative 
Names: Pastel 1, Dark 2, Dark 3, Set 2, Set 3, Warm, Cold, Harmonic,
       Dynamic

Type:  Sequential (single-hue) 
Names: Grays, Light Grays, Blues 2, Blues 3, Purples 2, Purples 3, Reds
       2, Reds 3, Greens 2, Greens 3, Oslo

Type:  Sequential (multi-hue) 
Names: Purple-Blue, Red-Purple, Red-Blue, Purple-Orange, Purple-Yellow,
       Blue-Yellow, Green-Yellow, Red-Yellow, Heat, Heat 2,
       Terrain, Terrain 2, Viridis, Plasma, Inferno, Dark Mint,
       Mint, BluGrn, Teal, TealGrn, Emrld, BluYl, ag_GrnYl, Peach,
       PinkYl, Burg, BurgYl, RedOr, OrYel, Purp, PurpOr, Sunset,
       Magenta, SunsetDark, ag_Sunset, BrwnYl, YlOrRd, YlOrBr,
       OrRd, Oranges, YlGn, YlGnBu, Reds, RdPu, PuRd, Purples,
       PuBuGn, PuBu, Greens, BuGn, GnBu, BuPu, Blues, Lajolla,
       Turku

Type:  Diverging 
Names: Blue-Red, Blue-Red 2, Blue-Red 3, Red-Green, Purple-Green,
       Purple-Brown, Green-Brown, Blue-Yellow 2, Blue-Yellow 3,
       Green-Orange, Cyan-Magenta, Tropic, Broc, Cork, Vik,
       Berlin, Lisbon, Tofino
\end{CodeOutput}
\end{CodeChunk}
To inspect the HCL parameter combinations for a specific palette simply
include the \texttt{palette} name where upper- vs.~lower-case, spaces,
etc.~are ignored for matching the label, i.e., \texttt{"set2"} matches
\texttt{"Set\ 2"}:
\begin{CodeChunk}
\begin{CodeInput}
R> hcl_palettes(palette = "set2")
\end{CodeInput}
\begin{CodeOutput}
HCL palette
Name: Set 2
Type: Qualitative
Parameter ranges:
 h1 h2 c1 c2 l1 l2 p1 p2 cmax fixup
  0 NA 60 NA 70 NA NA NA   NA  TRUE
\end{CodeOutput}
\end{CodeChunk}
To compute the actual color hex codes (representing sRGB coordinates)
based on these HCL parameters, the functions
\texttt{qualitative\_hcl()}, \texttt{sequential\_hcl()}, and
\texttt{diverging\_hcl()} can be used. Either all parameters can be
specified ``by hand'' through the HCL parameters, an entire palette can
be specified ``by name'', or the name-based specification can be
modified by a few HCL parameters. In case of the HCL parameters, either
a vector-based specification such as \texttt{h\ =\ c(0,\ 270)} or
individual parameters \texttt{h1\ =\ 0} and \texttt{h2\ =\ 270} can be
used.

The first three of the following commands lead to equivalent output. The
fourth command yields a modified set of colors (lighter due to a
luminance of 80 instead of 70).
\begin{CodeChunk}
\begin{CodeInput}
R> qualitative_hcl(4, h = c(0, 270), c = 60, l = 70)
\end{CodeInput}
\begin{CodeOutput}
[1] "#ED90A4" "#ABB150" "#00C1B2" "#ACA2EC"
\end{CodeOutput}
\begin{CodeInput}
R> qualitative_hcl(4, h1 = 0, h2 = 270, c1 = 60, l1 = 70)
\end{CodeInput}
\begin{CodeOutput}
[1] "#ED90A4" "#ABB150" "#00C1B2" "#ACA2EC"
\end{CodeOutput}
\begin{CodeInput}
R> qualitative_hcl(4, palette = "set2")
\end{CodeInput}
\begin{CodeOutput}
[1] "#ED90A4" "#ABB150" "#00C1B2" "#ACA2EC"
\end{CodeOutput}
\begin{CodeInput}
R> qualitative_hcl(4, palette = "set2", l = 80)
\end{CodeInput}
\begin{CodeOutput}
[1] "#FFACBF" "#C6CD70" "#32DDCD" "#C7BEFF"
\end{CodeOutput}
\end{CodeChunk}

\subsection{Qualitative palettes}\label{qualitative-palettes}

As suggested by \citet{color:Ihaka:2003} \texttt{qualitative\_hcl()}
distinguishes the underlying categories by a sequence of hues while
keeping both chroma and luminance constant, to give each color in the
resulting palette the same perceptual weight. Thus, \texttt{h} should be
a pair of hues (or equivalently \texttt{h1} and \texttt{h2} can be used)
with the starting and ending hue of the palette. Then, an equidistant
sequence between these hues is employed, by default spanning the full
color wheel (i.e., the full 360 degrees). Chroma \texttt{c} (or
equivalently \texttt{c1}) and luminance \texttt{l} (or equivalently
\texttt{l1}) are constants.

\begin{figure}[t!]
\centering
\includegraphics[width=0.8\textwidth]{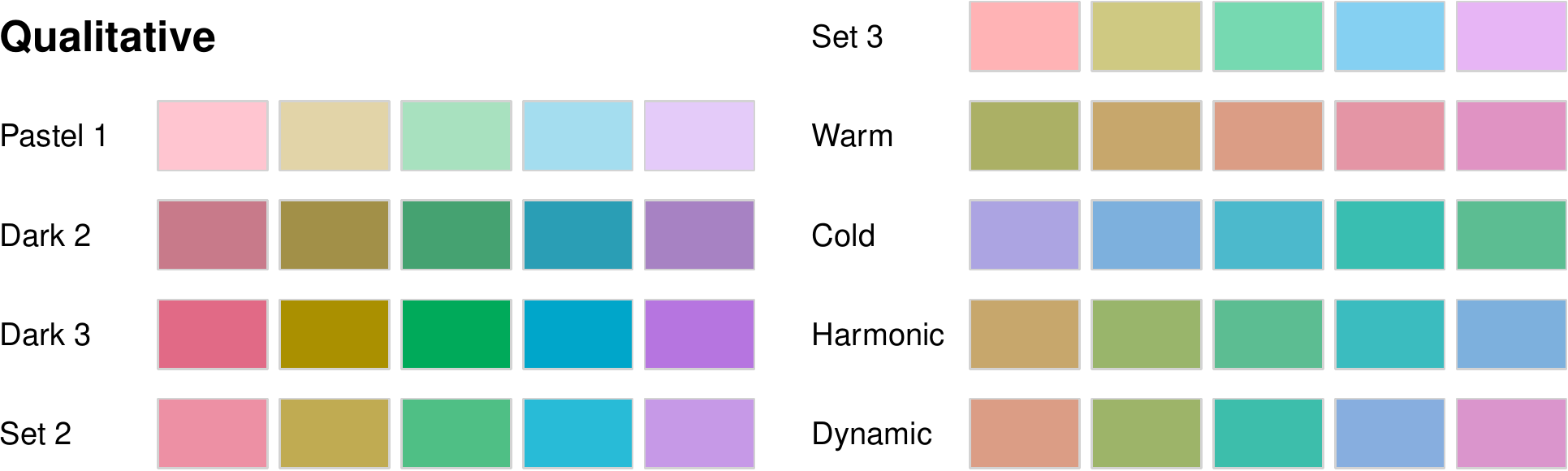} 
\caption[Prespecified qualitative HCL palettes available in \code{qualitative\_hcl()} in \pkg{colorspace}]{Prespecified qualitative HCL palettes available in \code{qualitative\_hcl()} in \pkg{colorspace}.}\label{fig:hcl-palettes-qualitative}
\end{figure}

Figure~\ref{fig:hcl-palettes-qualitative} shows the named palettes
available in the \texttt{qualitative\_hcl()} function. The first five
palettes are close to the \pkg{ColorBrewer.org} palettes of the same
name \citep{color:Harrower+Brewer:2003}. They employ different levels of
chroma and luminance and, by default, span the full hue range. The
remaining four palettes are taken from \citet{color:Ihaka:2003}. They
are based on the same chroma (= 50) and luminance (= 70) but the hue is
restricted to different intervals.
\begin{CodeChunk}
\begin{CodeInput}
R> hcl_palettes("qualitative", plot = TRUE, nrow = 5)
\end{CodeInput}
\end{CodeChunk}
When qualtitative palettes are employed for shading areas in statistical
displays (e.g., in bar plots, pie charts, or regions in maps), lighter
colors (with moderate chroma and high luminance) such as
\texttt{"Pastel\ 1"} or \texttt{"Set\ 3"} are typically less
distracting. By contrast, when coloring points or lines, more flashy
colors (with high chroma) are often required: On a white background a
moderate luminance as in \texttt{"Dark\ 2"} or \texttt{"Dark\ 3"}
usually works better while on a black/dark background the luminance
should be higher as in \texttt{"Set\ 2"}. Some examples with demo
graphics are provided in Section~\ref{sec:palette_visualization}.

\subsection{Sequential palettes (single-hue)}\label{sequential-palettes-single-hue}

\begin{figure}[t!]
\centering
\includegraphics[width=0.8\textwidth]{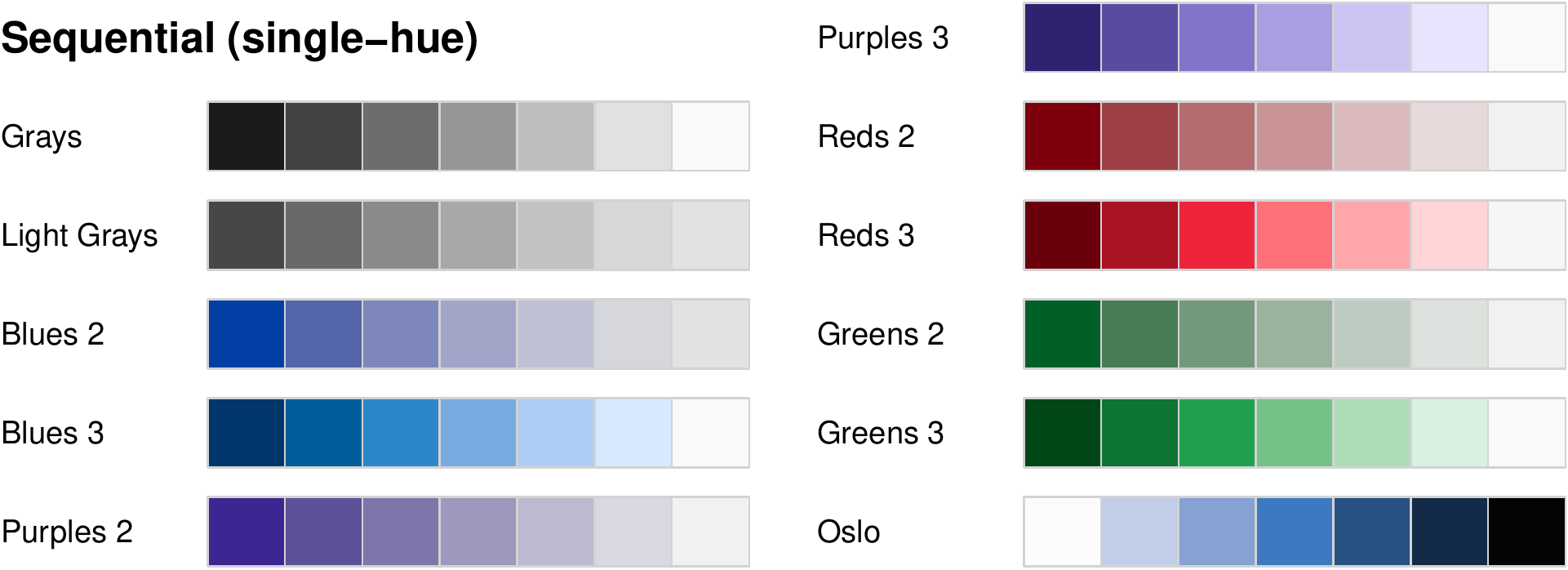} 
\caption[Prespecified sequential single-hue HCL palettes available in \code{sequential\_hcl()} in \pkg{colorspace}]{Prespecified sequential single-hue HCL palettes available in \code{sequential\_hcl()} in \pkg{colorspace}.}\label{fig:hcl-palettes-sequentials}
\end{figure}

As suggested by \citet{color:Zeileis+Hornik+Murrell:2009},
\texttt{sequential\_hcl()} codes the underlying numeric values by a
monotonic sequence of increasing (or decreasing) luminance. Thus, the
function's \texttt{l} argument should provide a vector of length 2 with
starting and ending luminance (equivalently, \texttt{l1} and \texttt{l2}
can be used). Without chroma (i.e., \texttt{c\ =\ 0}), this simply
corresponds to a grayscale palette like \texttt{gray.colors()}, see
\texttt{"Grays"} and \texttt{"Light\ Grays"} in
Figure~\ref{fig:hcl-palettes-sequentials}.

For adding chroma, a simple strategy would be to pick a single hue (via
\texttt{h} or \texttt{h1}) and then decrease chroma from some value
(\texttt{c} or \texttt{c1}) to zero (i.e., gray) along with increasing
luminance. For bringing out the extremes (a dark high-chroma color vs.~a
light gray) this is already very effective, see \texttt{"Blues\ 2"},
\texttt{"Purples\ 2"}, \texttt{"Reds\ 2"}, and \texttt{"Greens\ 2"}.

For distinguishing colors in the center of the palette, two strategies
can be employed: (a) Hue can be varied as well by specifying an interval
of hues in \texttt{h} (or beginning hue \texttt{h1} and ending hue
\texttt{h2}). More details are provided in the next section. (b) Instead
of a decreasing chroma, a triangular chroma trajectory can be employed
from \texttt{c1} over \texttt{cmax} to \texttt{c2} (equivalently
specified as a vector \texttt{c} of length 3). This yields high-chroma
colors in the middle of the palette that are more easily distinguished
from the dark and light extremes. See \texttt{"Blues\ 3"},
\texttt{"Purples\ 3"}, \texttt{"Reds\ 3"}, and \texttt{"Greens\ 3"} in
Figure~\ref{fig:hcl-palettes-sequentials}.

Instead of employing linear trajectories in the chroma or luminance
coordinates, some palettes employ a power transformation of the chroma
and/or luminance trajectory. Either a vector \texttt{power} of length 2
or separate \texttt{p1} (for chroma) and \texttt{p2} (for luminance) can
be specified. If the latter is missing, it defaults to the former.
\begin{CodeChunk}
\begin{CodeInput}
R> hcl_palettes("sequential (single-hue)", n = 7, plot = TRUE, nrow = 6)
\end{CodeInput}
\end{CodeChunk}
All except the last one are inspired by the \pkg{ColorBrewer.org}
palettes with the same base name \citep{color:Harrower+Brewer:2003} but
restricted to a single hue only. They are intended for a white/light
background. The last palette (\texttt{"Oslo"}) is taken from the
scientific color maps of \citet{color:Crameri:2018} and is intended for
a black/dark background and hence the order is reversed starting from a
light blue (not a light gray).

To distinguish many colors in a sequential palette it is important to
have a strong contrast on the luminance axis, possibly enhanced by an
accompanying pronounced variation in chroma. When only a few colors are
needed (e.g., for coding an ordinal categorical variable with few
levels) then a lower luminance contrast may suffice.

\subsection{Sequential palettes (multi-hue)}\label{sequential-palettes-multi-hue}

To not only bring out extreme colors in a sequential palette but also
better distinguish middle colors it is a common strategy to employ a
sequence of hues. Thus, the basis of such a palette is still a monotonic
luminance sequence as above (combined with a monotonic or triangular
chroma sequence). But rather than using a single hue, an interval of
hues in \texttt{h} (or beginning hue \texttt{h1} and ending hue
\texttt{h2}) can be specified.

\texttt{sequential\_hcl()} allows combined variations in hue (\texttt{h}
and \texttt{h1}/\texttt{h2}, respectively), chroma (\texttt{c} and
\texttt{c1}/\texttt{c2}/\texttt{cmax}, respectively), luminance
(\texttt{l} and \texttt{l1}/\texttt{l2}, respectively), and power
transformations for the chroma and luminance trajectories
(\texttt{power} and \texttt{p1}/\texttt{p2}, respectively). This yields
a broad variety of sequential palettes, including many that closely
match other well-known color palettes.
Figure~\ref{fig:hcl-palettes-sequentialm} shows all the named multi-hue
sequential palettes in \pkg{colorspace}:
\begin{CodeChunk}
\begin{CodeInput}
R> hcl_palettes("sequential (multi-hue)", n = 7, plot = TRUE)
\end{CodeInput}
%
\begin{figure}[t!]
\centering
\includegraphics[width=\textwidth]{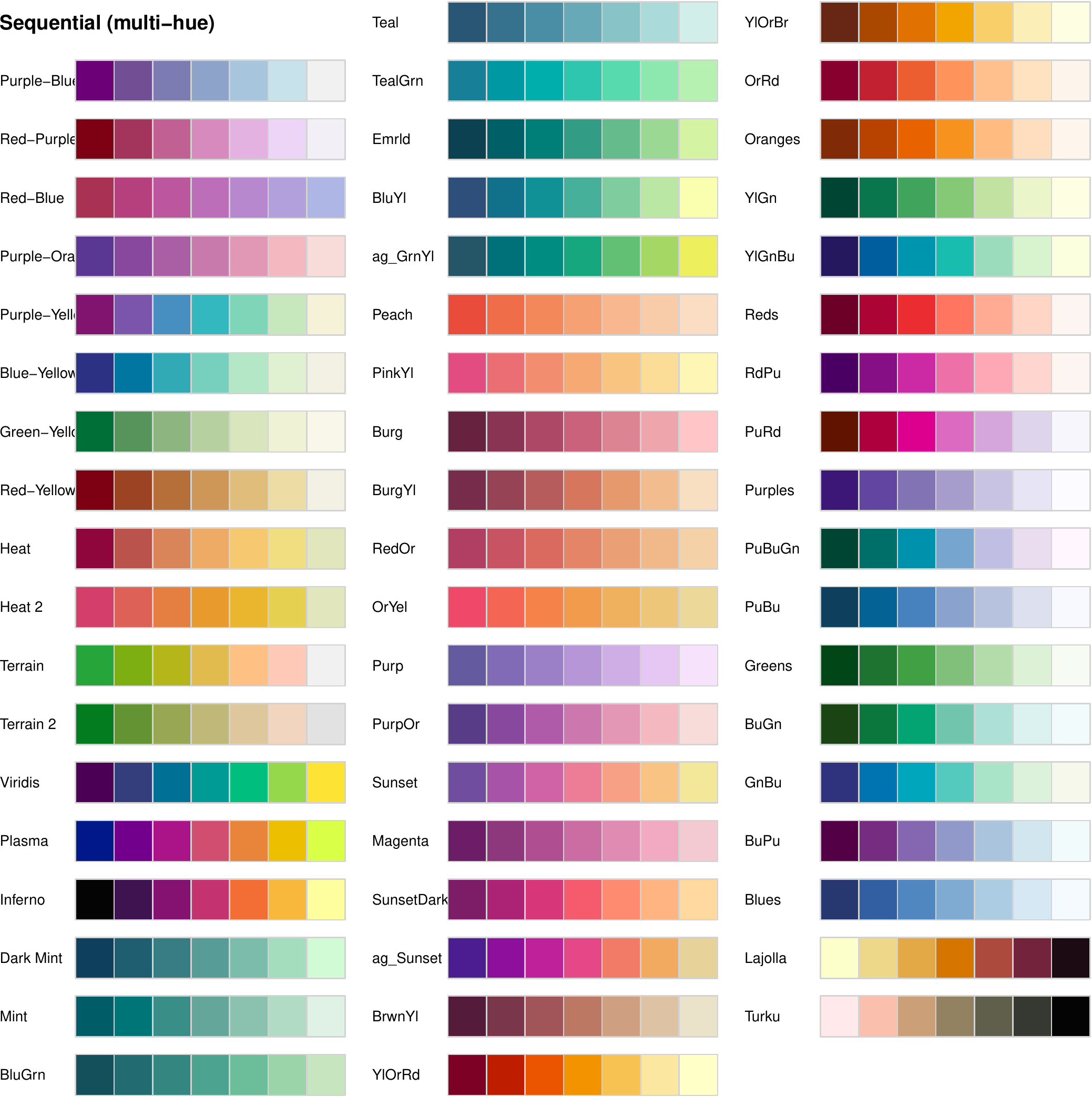} 
\caption[Prespecified sequential multi-hue HCL palettes available in \code{sequential\_hcl()} in \pkg{colorspace}]{Prespecified sequential multi-hue HCL palettes available in \code{sequential\_hcl()} in \pkg{colorspace}.}\label{fig:hcl-palettes-sequentialm}
\end{figure}
%
\end{CodeChunk}
\begin{itemize}
\tightlist
\item
  \texttt{"Purple-Blue"} to \texttt{"Terrain\ 2"} are various palettes
  created during the development of \pkg{colorspace}, e.g., by
  \citet{color:Zeileis+Hornik+Murrell:2009} or
  \citet{color:Stauffer+Mayr+Dabernig:2015} among others.
\item
  \texttt{"Viridis"} to \texttt{"Inferno"} closely match the palettes
  that \citet{color:Smith+VanDerWalt:2015} developed for
  \pkg{matplotlib} and that gained popularity recently.
\item
  \texttt{"Dark\ Mint"} to \texttt{"BrwnYl"} closely match palettes
  provided in \pkg{CARTO} \citep{color:CARTO}.
\item
  \texttt{"YlOrRd"} to \texttt{"Blues"} closely match
  \pkg{ColorBrewer.org} palettes \citep{color:Harrower+Brewer:2003}.
\item
  \texttt{"Lajolla"} and \texttt{"Turku"} closely match the scientific
  color maps of the same name by \citet{color:Crameri:2018} and are
  intended for a black/dark background.
\end{itemize}
Note that the palettes differ substantially in the amount of chroma and
luminance contrasts, respectively. For example, many palettes go from a
dark high-chroma color to a neutral low-chroma color (e.g.,
\texttt{"Reds"}, \texttt{"Purples"}, \texttt{"Greens"},
\texttt{"Blues"}) or even light gray (e.g., \texttt{"Purple-Blue"}). But
some palettes also employ relatively high chroma throughout the palette
(e.g., the viridis and many \pkg{CARTO} palettes). To emphasize the
extremes the former strategy is typically more suitable while the latter
works better if all values along the sequence should receive some more
perceptual weight.

\subsection{Diverging palettes}\label{diverging-palettes}

\texttt{diverging\_hcl()} codes the underlying numeric values by a
triangular luminance sequence with different hues in the left and in the
right ``arm'' of the palette. Thus, it can be seen as a combination of
two sequential palettes with some restrictions: (a) a single hue is used
for each arm of the palette, (b) chroma and luminance trajectory are
balanced between the two arms, (c) the neutral central value has zero
chroma. To specify such a palette a vector of two hues \texttt{h} (or
equivalently \texttt{h1} and \texttt{h2}), either a single chroma value
\texttt{c} (or \texttt{c1}) or a vector of two chroma values \texttt{c}
(or \texttt{c1} and \texttt{cmax}), a vector of two luminances
\texttt{l} (or \texttt{l1} and \texttt{l2}), and power parameter(s)
\texttt{power} (or \texttt{p1} and \texttt{p2}) are used. For more
flexible diverging palettes without the restrictrictions above (and
consequently more parameters) see the \texttt{divergingx\_hcl()}
palettes introduced below.

\begin{figure}[t!]
\centering
\includegraphics[width=0.8\textwidth]{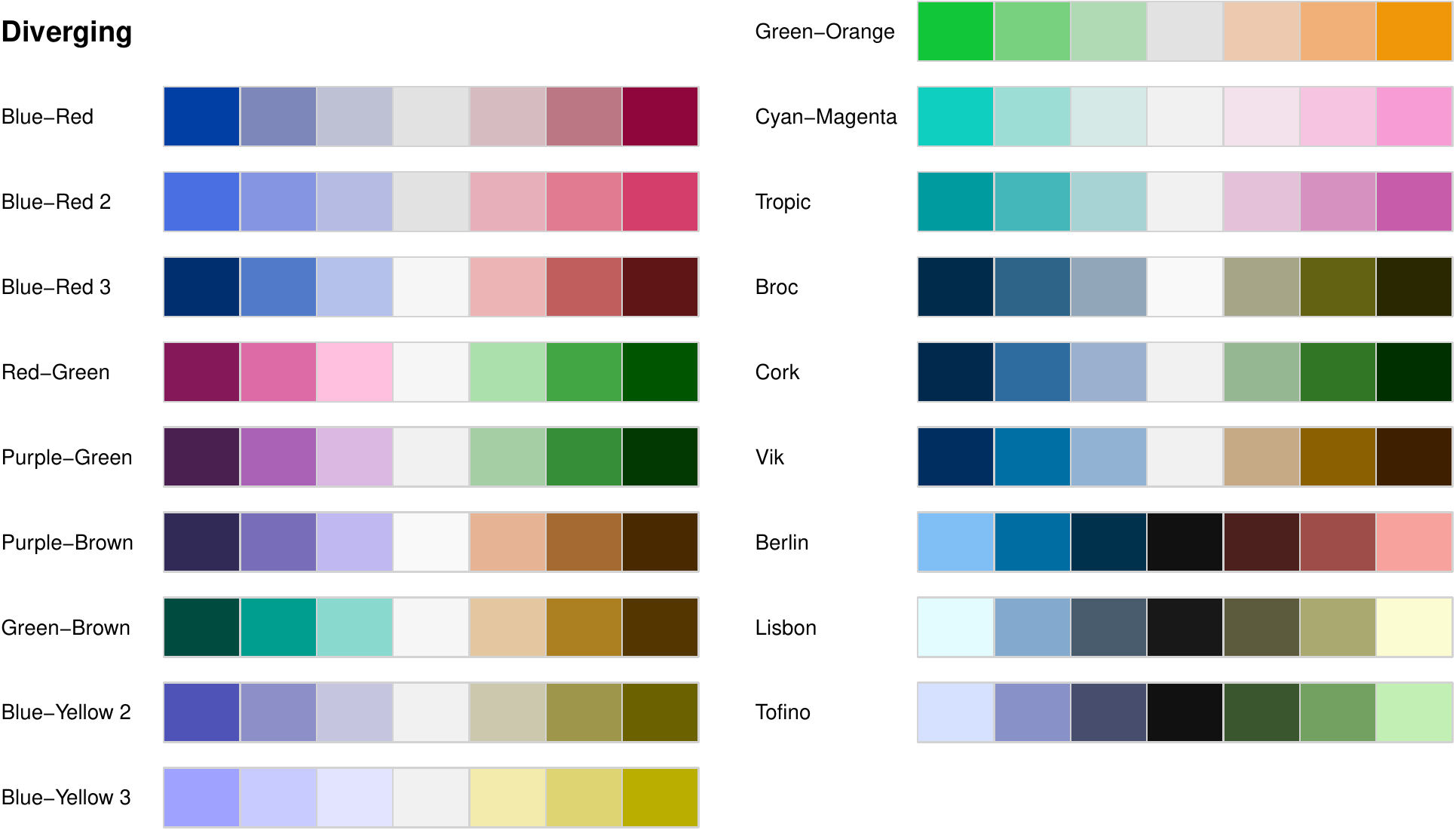} 
\caption[Prespecified diverging HCL palettes available in \code{diverging\_hcl()} in \pkg{colorspace}]{Prespecified diverging HCL palettes available in \code{diverging\_hcl()} in \pkg{colorspace}.}\label{fig:hcl-palettes-diverging}
\end{figure}

Figure~\ref{fig:hcl-palettes-diverging} shows all such diverging
palettes that have been named in \pkg{colorspace}:
\begin{CodeChunk}
\begin{CodeInput}
R> hcl_palettes("diverging", n = 7, plot = TRUE, nrow = 10)
\end{CodeInput}
\end{CodeChunk}
\begin{itemize}
\tightlist
\item
  \texttt{"Blue-Red"} to \texttt{"Cyan-Magenta"} have been developed for
  \pkg{colorspace} starting from
  \citet{color:Zeileis+Hornik+Murrell:2009}, taking inspiration from
  various other palettes, including more balanced and simplified
  versions of several \pkg{ColorBrewer.org} palettes
  \citep{color:Harrower+Brewer:2003}.
\item
  \texttt{"Tropic"} closely matches the palette of the same name from
  \pkg{CARTO} \citep{color:CARTO}.
\item
  \texttt{"Broc"} to \texttt{"Vik"} and \texttt{"Berlin"} to
  \texttt{"Tofino"} closely match the scientific color maps of the same
  name by \citet{color:Crameri:2018}, where the first three are intended
  for a white/light background and the other three for a black/dark
  background.
\end{itemize}
When choosing a particular palette for a display similar considerations
apply as for the sequential palettes. Thus, large luminance differences
are important when many colors are used while smaller luminance
contrasts may suffice for palettes with fewer colors etc.

\subsection{Construction details}\label{construction-details}

Table~\ref{tab:hcl} summarizes which types of trajectories
(\emph{constant}, \emph{linear}, \emph{triangular}) are used for the
three HCL coordinates (hue \emph{H}, chroma \emph{C}, luminance
\emph{L}) to construct the different types of palettes
(\emph{qualitative}, \emph{sequential}, and \emph{diverging}).

\begin{table}[t!]
\centering
\begin{tabular}{llll}
\hline
Type          & $H$                               & $C$                        & $L$              \\ \hline
Qualitative   & Linear                    & Constant                   & Constant     \\
Sequential    & Constant (= single-hue) \emph{or} & Linear (+ power) \emph{or} & Linear (+ power) \\
          & Linear (= multi-hue)              & Triangular (+ power)       &          \\
Diverging     & Constant ($2\times$)              & Linear (+ power) \emph{or} & Linear (+ power) \\
          &                       & Triangular (+ power)       &          \\
\hline
\end{tabular}
\caption{\label{tab:hcl} Types of trajectories used for the HCL coordinates to construct qualitative,
sequential, and diverging palettes, see Equations~\ref{eq:constant}--\ref{eq:triangular}.}
\end{table}

As emphasized in Figure~\ref{fig:hcl-palettes-principles}, the luminance
is probably the most important property for defining the type of
palette. It is constant for qualitative palettes, monotonic for
sequential palettes (linear or a power transformation), or uses two
monotonic trajectories (linear or a power transformation) diverging from
the same neutral value.

\begin{figure}[t!]
\centering
\includegraphics[width=\textwidth]{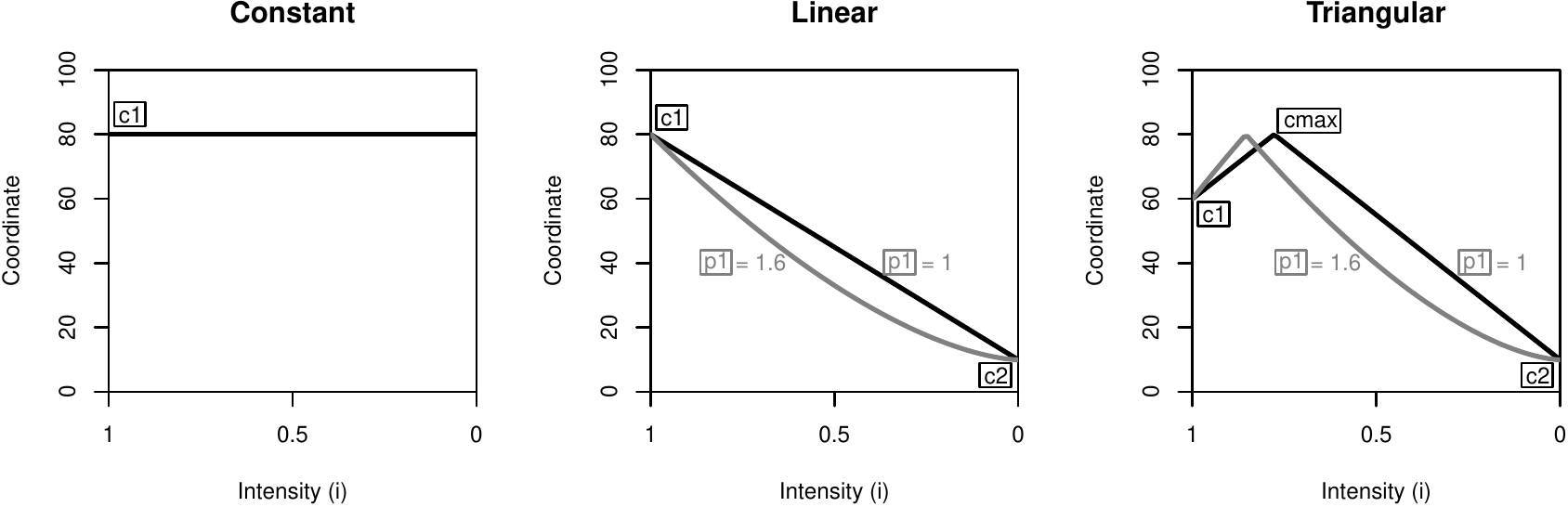} 
\caption{Types of trajectories to construct HCL color palettes, exemplified for the chroma coordinates, see Equations~\ref{eq:constant}--\ref{eq:triangular}.}\label{fig:hcl-trajectories}
\end{figure}

The trajectories for the hue are also rather intuitive and
straightforward for the three different types of palettes (constant
vs.~linear). However, the chroma trajectories are probably most
complicated and least obvious from the examples above. Hence, the exact
mathematical equations underlying the chroma trajectories are given in
the following (i.e., using the parameters \texttt{c1}, \texttt{c2},
\texttt{cmax}, and \texttt{p1}, respectively) and are depicted in
Figure~\ref{fig:hcl-trajectories}. Analogous equations apply for the
other two coordinates.

The trajectories are functions of the \emph{intensity} \(i \in [0, 1]\)
where \(1\) corresponds to the full intensity:
\begin{align}
\text{Constant: }   & c_1 \label{eq:constant} \\[0.2cm]
\text{Linear: }     & c_2 - (c_2 - c_1) \cdot i \label{eq:linear} \\[0.2cm]
\text{Triangular: } & \left\{ \begin{array}{lcl}
                      c_2 - (c_2 - c_{\max}) \cdot \frac{i}{j}            & \text{if } i & \le j \\
                      c_{\max} - (c_{\max} - c_1) \cdot \frac{i - j}{1 - j} &              & >   j
                     \end{array} \right. \label{eq:triangular}
\end{align}
where \(j\) is the intensity at which \(c_{\max}\) is assumed. It is
constructed such that the slope to the left is minus the slope to the
right of \(j\):
\[
j = \left(1 + \frac{|c_{\max} - c_1|}{|c_{\max} - c_2|} \right)^{-1}
\]
Instead of using a linear intensity \(i\) going from \(1\) to \(0\), one
can replace \(i\) with \(i^{p_1}\) in
Equations~\ref{eq:constant}--\ref{eq:triangular}. This then leads to
power-transformed curves that add or remove chroma more slowly or more
quickly depending on whether the power parameter \(p_1\) is \(< 1\) or
\(> 1\).

The three types of trajectories are also depicted in
Figure~\ref{fig:hcl-trajectories}. Note that full intensity \(i = 1\) is
on the left and zero intensity \(i = 0\) on the right of each panel. The
concrete parameters are:
\begin{itemize}
\tightlist
\item
  Constant: \texttt{c1\ =\ 80}.
\item
  Linear: \texttt{c1\ =\ 80}, \texttt{c2\ =\ 10}, \texttt{p1\ =\ 1}
  (solid) vs.~\texttt{p1\ =\ 1.6} (dashed).
\item
  Triangular: \texttt{c1\ =\ 60}, \texttt{cmax\ =\ 80},
  \texttt{c2\ =\ 10}, \texttt{p1\ =\ 1} (solid) vs.~\texttt{p1\ =\ 1.6}
  (dashed).
\end{itemize}
Further discussion of these trajectories and how they can be visualized
and assessed for a given color palette is provided in
Section~\ref{sec:palette_visualization}.

\subsection{Registering your own palettes}\label{registering-your-own-palettes}

The \texttt{hcl\_palettes()} already come with a wide range of
predefined palettes to which customizations can be easily added.
However, it might also be convenient to register a custom palette so
that it can subsequently be reused with a new dedicated name. This is
supported by adding a \texttt{register\ =\ "..."} argument once to a
call to \texttt{qualitative\_hcl()}, \texttt{sequential\_hcl()}, or
\texttt{diverging\_hcl()}:
\begin{CodeChunk}
\begin{CodeInput}
R> qualitative_hcl(3, palette = "set2", l = 80, register = "myset")
\end{CodeInput}
\end{CodeChunk}
The new palette is then included in \texttt{hcl\_palettes()}:
\begin{CodeChunk}
\begin{CodeInput}
R> hcl_palettes("Qualitative")
\end{CodeInput}
\begin{CodeOutput}
HCL palettes

Type:  Qualitative 
Names: Pastel 1, Dark 2, Dark 3, Set 2, Set 3, Warm, Cold, Harmonic,
       Dynamic, myset
\end{CodeOutput}
\end{CodeChunk}
And can be used subsequently in \texttt{qualitative\_hcl()} as well as
the qualitative \pkg{ggplot2} color scales (see
Section~\ref{sec:ggplot2}), e.g.,
\begin{CodeChunk}
\begin{CodeInput}
R> qualitative_hcl(4, palette = "myset")
\end{CodeInput}
\begin{CodeOutput}
[1] "#FFACBF" "#C6CD70" "#32DDCD" "#C7BEFF"
\end{CodeOutput}
\end{CodeChunk}
Remarks:
\begin{itemize}
\tightlist
\item
  The number of colors in the palette that was used during registration
  is not actually stored and can be modified subsequently. The same
  holds for arguments \texttt{alpha} and \texttt{rev}.
\item
  When registering a new palette with an old name that was already
  available previously, the old palette gets overwritten. We recommend
  not to overwrite the palettes that are predefined in the package
  (albeit it is technically possible).
\item
  The registration of a palette is only stored for the current session.
  When \proglang{R} is restarted and/or the \pkg{colorspace} package
  reloaded, only the predefined palettes from the package are available.
  Thus, to make a palette permanently available a registration
  \proglang{R} code like
  \texttt{colorspace::qualitative\_hcl(3,\ palette\ =\ "set2",\ l\ =\ 80,\ register\ =\ "myset")}
  can be placed in your \texttt{.Rprofile} or similar startup scripts.
\end{itemize}

\subsection{Flexible diverging palettes}\label{sec:divergingx}

The \texttt{divergingx\_hcl()} function provides more flexible diverging
palettes by simply calling \texttt{sequential\_hcl()} twice with
prespecified sets of hue, chroma, and luminance parameters. Thus, it
does not pose any restrictions that the two ``arms'' of the palette need
to be balanced and also allows to go through a non-gray neutral color
(typically light yellow). Consequently, the chroma/luminance paths can
be rather unbalanced.

\begin{figure}[t!]
\centering
\includegraphics[width=0.8\textwidth]{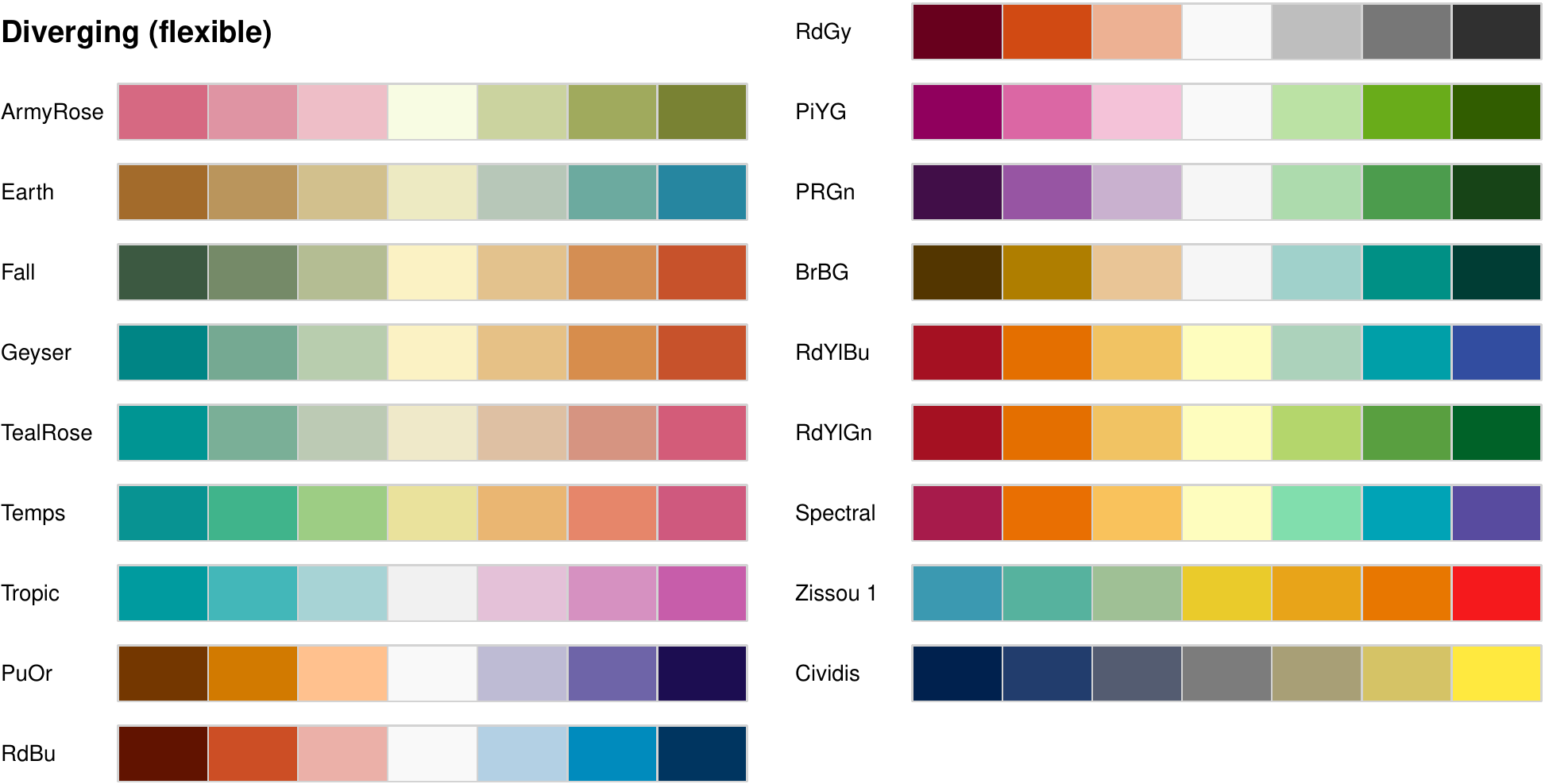} 
\caption[Prespecified flexible diverging HCL palettes available in \code{divergingx\_hcl()} in \pkg{colorspace}]{Prespecified flexible diverging HCL palettes available in \code{divergingx\_hcl()} in \pkg{colorspace}.}\label{fig:divergingx-palettes}
\end{figure}

Figure~\ref{fig:divergingx-palettes} shows all such flexible diverging
palettes that have been named in \pkg{colorspace}:
\begin{CodeChunk}
\begin{CodeInput}
R> divergingx_palettes(n = 7, plot = TRUE, nrow = 10)
\end{CodeInput}
\end{CodeChunk}
\begin{itemize}
\tightlist
\item
  \texttt{"ArmyRose"} to \texttt{"Tropic"} closely match the palettes of
  the same name from \pkg{CARTO} \citep{color:CARTO}.
\item
  \texttt{"PuOr"} to \texttt{"Spectral"} closely match the palettes of
  the same name from \pkg{ColorBrewer.org}
  \citep{color:Harrower+Brewer:2003}.
\item
  \texttt{"Zissou\ 1"} closely matches the palette of the same name from
  \pkg{wesanderson} \citep{color:wesanderson}.
\item
  \texttt{"Cividis"} closely matches the palette of the same name from
  the \pkg{viridis} family \citep{color:viridis}. Note that despite
  having two ``arms'' with blue vs.~yellow colors and a low-chroma
  center color, this is probably better classified as a sequential
  palette due to the monotonic chroma going from dark to light. (See
  Section~\ref{sec:approximations} for more details.)
\end{itemize}
Typically, the more restricted \texttt{diverging\_hcl()} palettes should
be preferred because they are more balanced. However, by being able to
go through light yellow as the neutral color warmer diverging palettes
are available.

\subsection{Approximating palettes from other packages}\label{sec:approximations}

The flexible specification of HCL-based color palettes in
\pkg{colorspace} allows to closely approximate color palettes from
various other packages:
\begin{itemize}
\tightlist
\item
  \pkg{ColorBrewer.org} \citep{color:Harrower+Brewer:2003} as provided
  by \proglang{R} package \pkg{RColorBrewer} \citep{color:RColorBrewer}.
  See \texttt{demo("brewer",\ package\ =\ "colorspace")}.
\item
  \pkg{CARTO} colors \citep{color:CARTO} as provided by \proglang{R}
  package \pkg{rcartocolor} \citep{color:rcartocolor}. See
  \texttt{demo("carto",\ package\ =\ "colorspace")}.
\item
  The viridis palettes of \citet{color:Smith+VanDerWalt:2015} developed
  for \pkg{matplotlib}, as provided by \proglang{R} package
  \pkg{viridis} \citep{color:viridis}. See
  \texttt{demo("viridis", package\ =}\linebreak \texttt{"colorspace")}.
\item
  The scientific color maps of \citet{color:Crameri:2018} as provided by
  \proglang{R} package \pkg{scico} \citep{color:scico}. See
  \texttt{demo("scico", package\ =\ "colorspace")}.
\end{itemize}
The graphics resulting from the demos can also be viewed online at
\url{http://colorspace.R-Forge.R-project.org/articles/approximations.html}.

\begin{figure}[t!]
\centering
\includegraphics[width=0.49\textwidth]{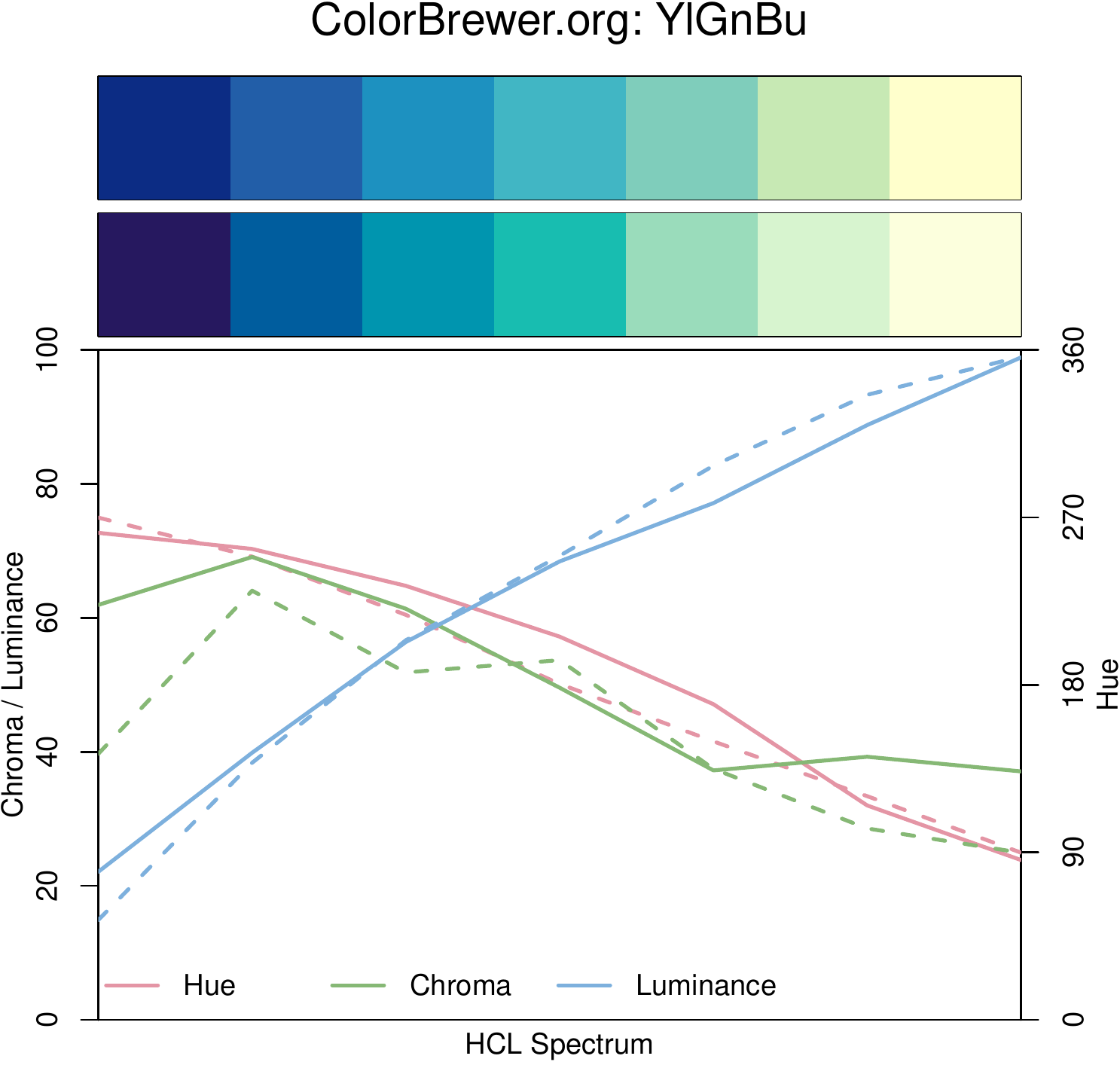} \includegraphics[width=0.49\textwidth]{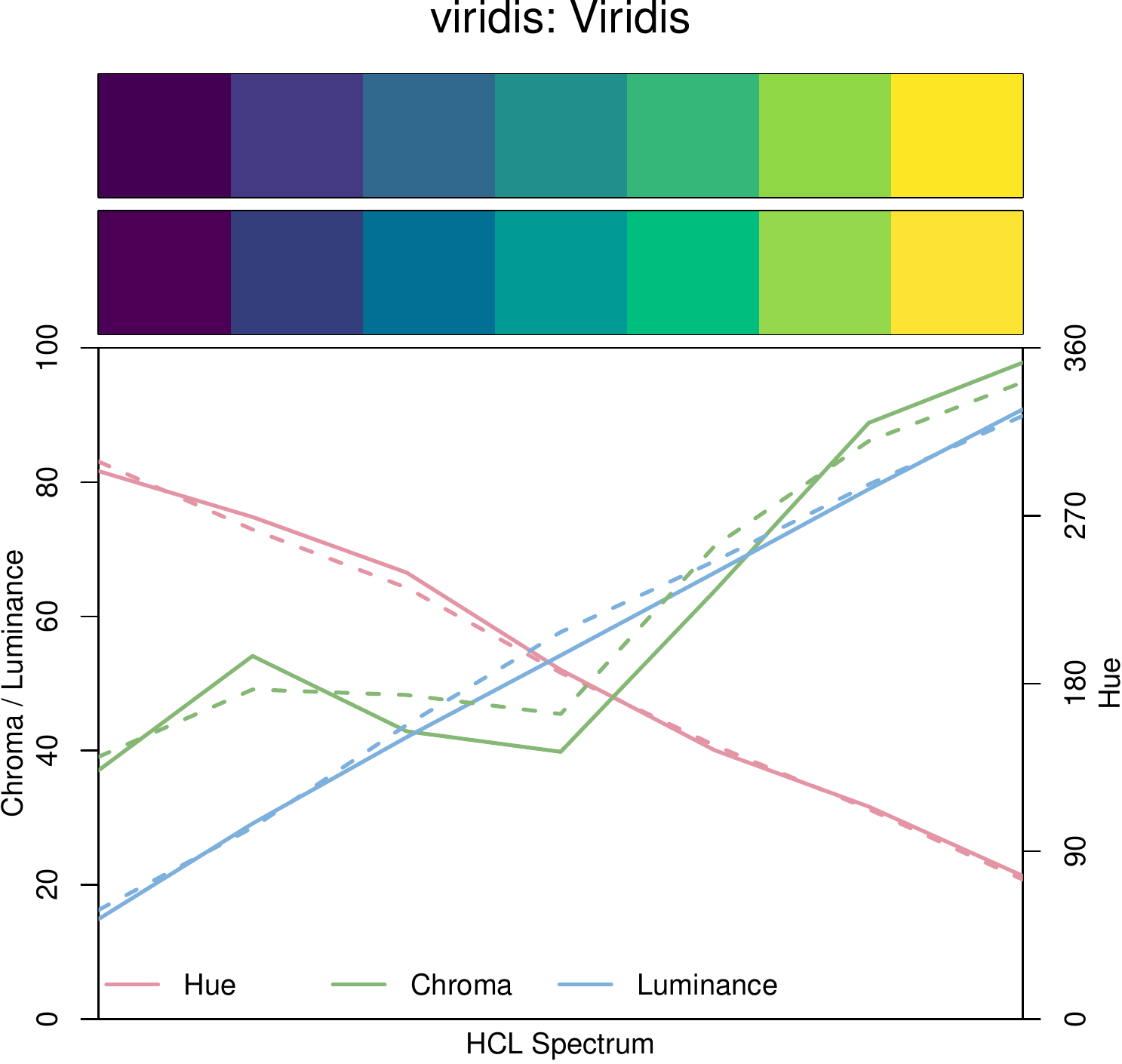} \includegraphics[width=0.49\textwidth]{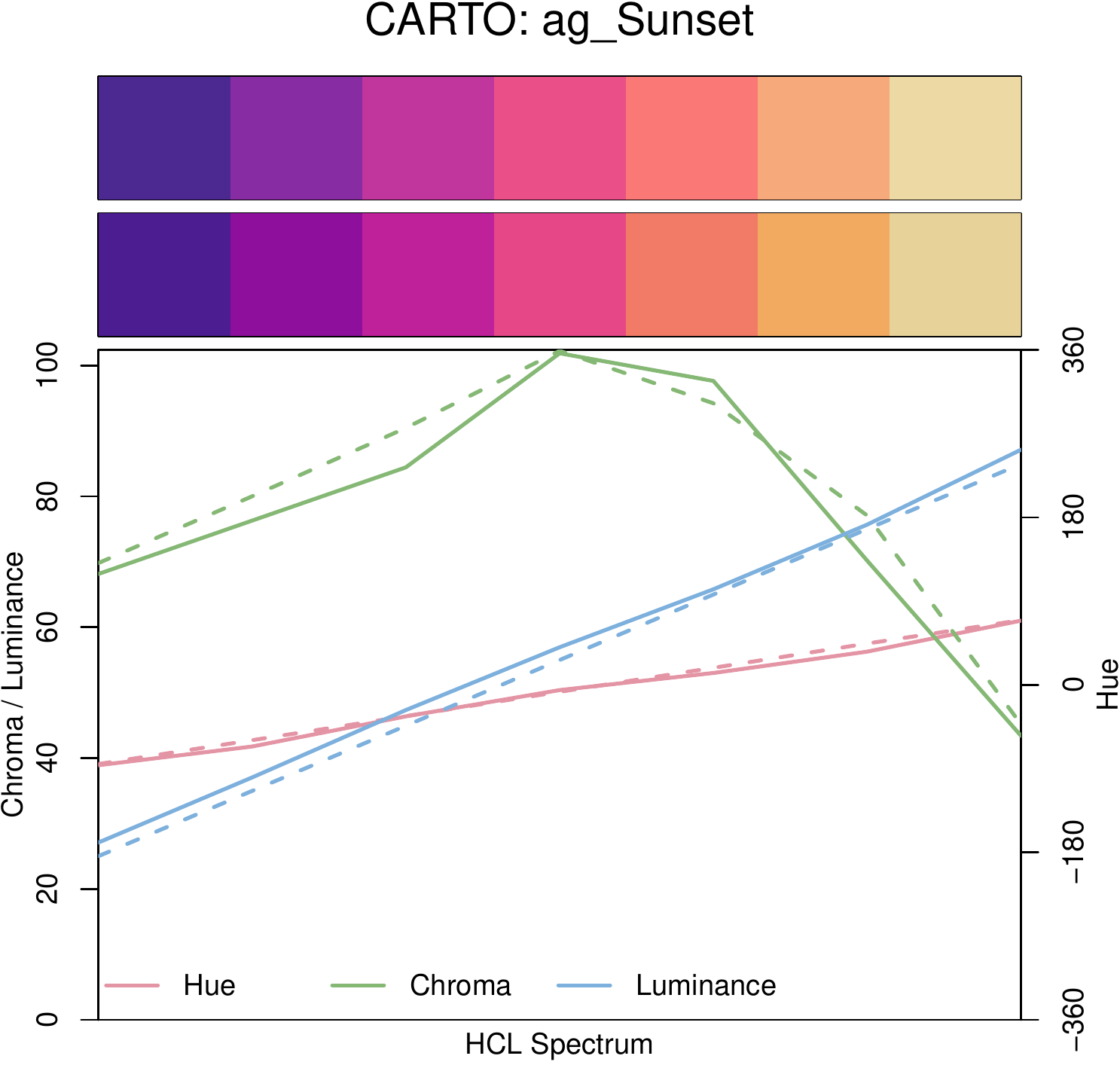} \includegraphics[width=0.49\textwidth]{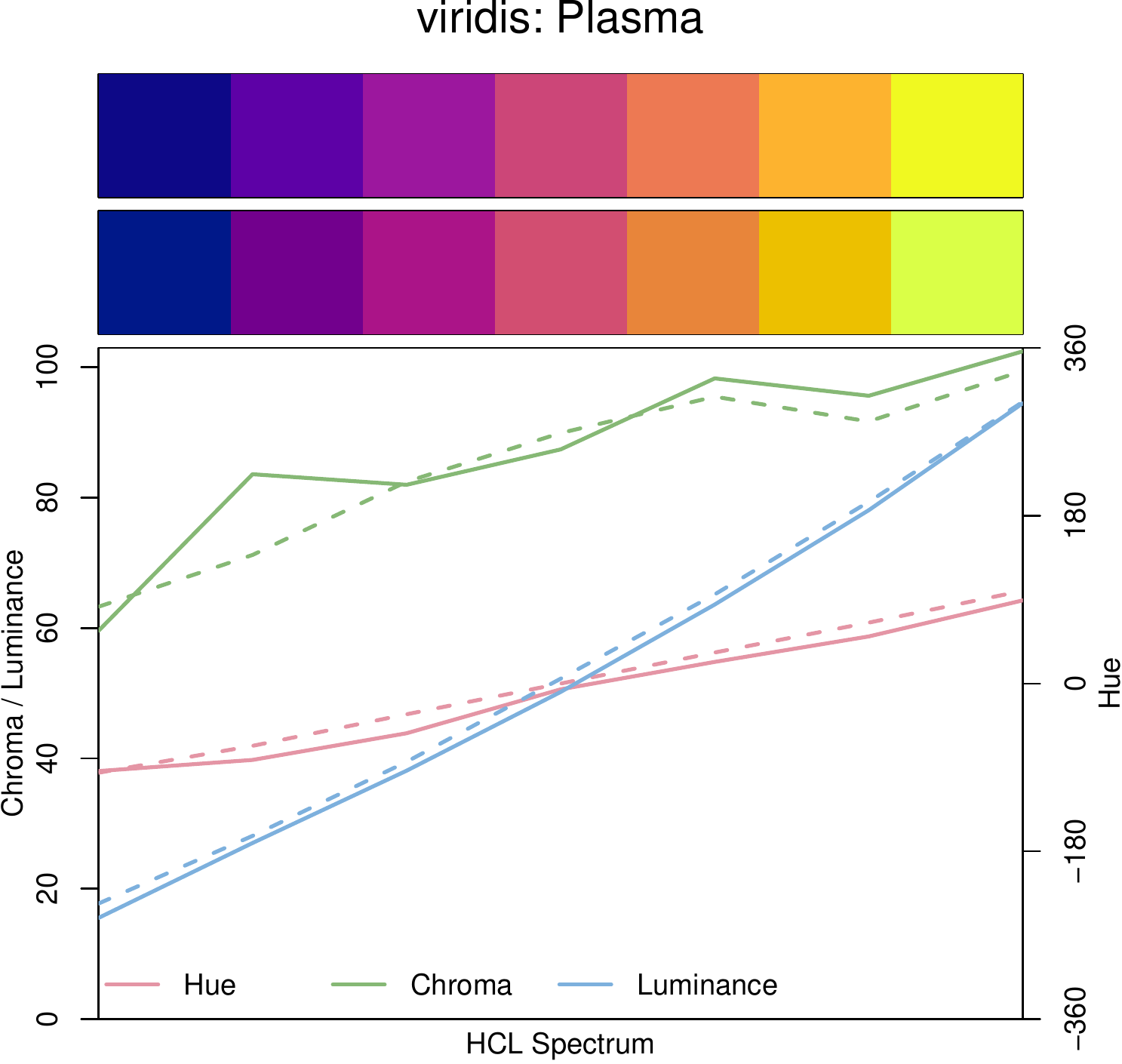} 
\caption{HCL spectrum of four palettes taken from \pkg{ColorBrewer.org}, \pkg{CARTO}, and \pkg{viridis} (top swatches, solid lines) along with their HCL-based approximations (bottom swatches, dashed lines).}\label{fig:brewer-carto-viridis}
\end{figure}

Figure~\ref{fig:brewer-carto-viridis} shows a selection of such
approximations using \texttt{specplot()} (see also
Section~\ref{sec:specplot}) for two blue/green/yellow palettes
(\texttt{RColorBrewer::brewer.pal(7,\ "YlGnBu")} and\linebreak
\texttt{viridis::viridis(7)}) and two purple/red/yellow
(\texttt{rcartocolor::carto\_pal(7,}\linebreak \texttt{"ag\_Sunset")} and
\texttt{viridis::plasma(7)}). Each panel compares the hue, chroma, and
luminance trajectories of the original palettes (top swatches, solid
lines) and their HCL-based approximations (bottom swatches, dashed
lines). The palettes are not identical but very close for most colors.
Note that also the chroma trajectories from the HCL palettes (green
dashed lines) have some kinks which are due to fixing HCL coordinates at
the boundaries of admissible RGB colors.

Furthermore, Figure~\ref{fig:brewer-carto-viridis} illustrates what sets
the viridis palettes apart from other sequential palettes. While the hue
and luminance trajectories of \texttt{"Viridis"} and \texttt{"YlGnBu"}
are very similar, the chroma trajectories differ: While lighter colors
(with high luminance) have low chroma for \texttt{"YlGnBu"}, they have
increasing chroma for \texttt{"Viridis"}. Similarly,
\texttt{"ag\_Sunset"} and \texttt{"Plasma"} have similar hue and
luminance trajectories but different chroma trajectories. The result is
that the viridis palettes have rather high chroma throughout which does
not work as well for sequential palettes on a white/light background as
all shaded areas convey high ``intensity''. However, they work better on
a dark/black background (see Figure~\ref{fig:demoplot-dark}). Also, they
might be a reasonable alternative for qualitative palettes when
grayscale printing should also work.

\begin{figure}[t!]
\centering
\includegraphics[width=0.49\textwidth]{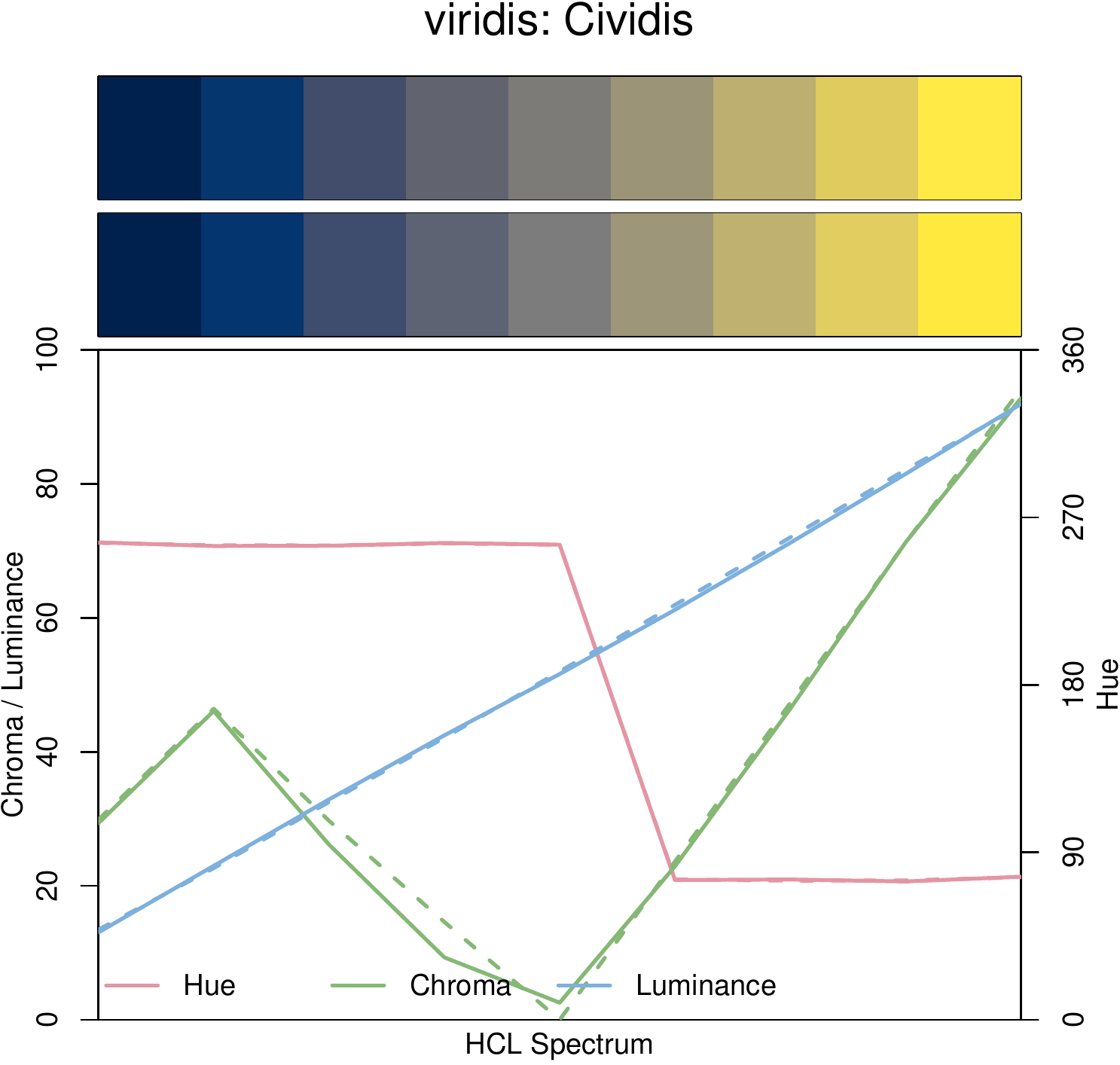} 
\caption{HCL spectrum of \texttt{viridis::cividis} (top swatch, solid lines) along with an HCL-based approximation (bottom swatch, dashed lines).}\label{fig:cividis}
\end{figure}

Another somewhat nonstandard palette from the viridis family is the
cividis palette based on blue and yellow hues and hence safe for
red-green deficient viewers. Figure~\ref{fig:cividis} shows the
corresponding \texttt{specplot()} along with an HCL-based approximation.
What is unusual about this palette: The hue and chroma trajectories
would suggest a diverging palette, as there are two ``arms'' wth
different hues and a zero-chroma point in the center. However, the
luminance trajectory clearly indicates a sequential palette as colors go
monotonically from dark to light. Due to this unusual mixture the
palette cannot be composed using the trajectories from
Table~\ref{tab:hcl}.

However, the tools in \pkg{colorspace} can still be employed to easily
reconstruct the palette. One strategy would be to set up the
trajectories manually, using a linear luminance, piecewise linear
chroma, and piecewise constant hue:
\begin{CodeChunk}
\begin{CodeInput}
R> cividis_hcl <- function(n) {
+    i <- seq(1, 0, length.out = n)
+    hex(polarLUV(
+      L = 92 - (92 - 13) * i,
+      C = approx(c(1, 0.9, 0.5, 0), c(30, 50, 0, 95), xout = i)$y,
+      H = c(255, 75)[1 + (i < 0.5)]
+    ), fix = TRUE)
+  }
\end{CodeInput}
\end{CodeChunk}
Instead of constructing the hex code from the HCL coordinates via
\texttt{hex(polarLUV(L,\ C,\ H))} from \pkg{colorspace}, the base
\proglang{R} function \texttt{hcl(H,\ C,\ L)} from \pkg{grDevices} could
also be used.

In addition to manually setting up a dedicated function
\texttt{cividis\_hcl()}, it is possible to approximate the palette using
\texttt{divergingx\_hcl()} (see Section~\ref{sec:divergingx}), e.g.,
\begin{CodeChunk}
\begin{CodeInput}
R> divergingx_hcl(n,
R>   h1 = 255, h2 = NA, h3 = 75,
R>   c1 = 30, cmax1 = 47, c2 = 0, c3 = 95,
R>   l1 = 13, l2 = 52, l3 = 92,
R>   p1 = 1.1, p3 = 1.0
R> )
\end{CodeInput}
\end{CodeChunk}
This uses a slight power transformation with \texttt{p1\ =\ 1.1} in the
blue arm of the palette but otherwise essentially corresponds to what
\texttt{cividis\_hcl()} does. For convenience
\texttt{divergingx\_hcl(n,\ palette\ =\ "Cividis")} is preregistered
using the above parameters.

\subsection[HCL (and HSV) color palettes corresponding to base R palettes]{HCL (and HSV) color palettes corresponding to base \proglang{R} palettes}\label{hcl-and-hsv-color-palettes-corresponding-to-base-r-palettes}

To facilitate switching from base \proglang{R} palette functions to the
HCL-based palettes above, \pkg{colorspace} provides a few convenience
interfaces:
\begin{itemize}
\tightlist
\item
  \texttt{rainbow\_hcl()}: Convenience interface to
  \texttt{qualitative\_hcl()} for a HCL-based ``rainbow'' palette to
  replace the (in)famous \texttt{rainbow()} palette.
\item
  \texttt{heat\_hcl()}: Convenience interface to
  \texttt{sequential\_hcl()} with default parameters chosen to generate
  more balanced heat colors than the basic \texttt{heat.colors()}
  function.
\item
  \texttt{terrain\_hcl()}: Convenience interface to
  \texttt{sequential\_hcl()} with default parameters chosen to generate
  more balanced terrain colors than the basic \texttt{terrain.colors()}
  function.
\item
  \texttt{diverging\_hsv()}: Diverging palettes generated in HSV space
  rather than HCL space as in \texttt{diverging\_hcl()}. This is
  provided for didactic purposes to contrast the more balanced HCL
  palettes with the more flashy and unbalanced HSV palettes.
\end{itemize}

\section{Palette visualization and assessment}\label{sec:palette_visualization}

The \pkg{colorspace} package provides several visualization functions
for depicting one or more color palettes and their underlying
properties. Color palettes can be visualized by:
\begin{itemize}
\tightlist
\item
  \texttt{swatchplot()}: Color swatches.
\item
  \texttt{specplot()}: Spectrum of HCL and/or RGB trajectories.
\item
  \texttt{hclplot()}: Trajectories in 2-dimensional HCL space
  projections.
\item
  \texttt{demoplot()}: Illustrations of typical (and simplified)
  statistical graphics.
\end{itemize}

\subsection{Color swatches}\label{sec:swatchplot}

The function \texttt{swatchplot()} is a convenience function for
displaying collections of palettes that can be specified as lists or
matrices of hex color codes. Essentially, it is just a call to the base
graphics \texttt{rect()} function but with heuristics for choosing
default labels, margins, spacings, borders, etc. These heuristics are
selected to work well for \texttt{hcl\_palettes()} and might need
further tweaking in future versions of the package. Thus,
Figures~\ref{fig:hcl-properties}--\ref{fig:hcl-palettes} as well as
Figures~\ref{fig:hcl-palettes-principles}--\ref{fig:hcl-palettes-diverging}
all use \texttt{swatchplot()} internally. For a simple stand-alone
illustration consider:
\texttt{swatchplot("Palette"\ =\ sequential\_hcl(5))}.

\begin{figure}[t!]
\centering
\includegraphics[width=\textwidth]{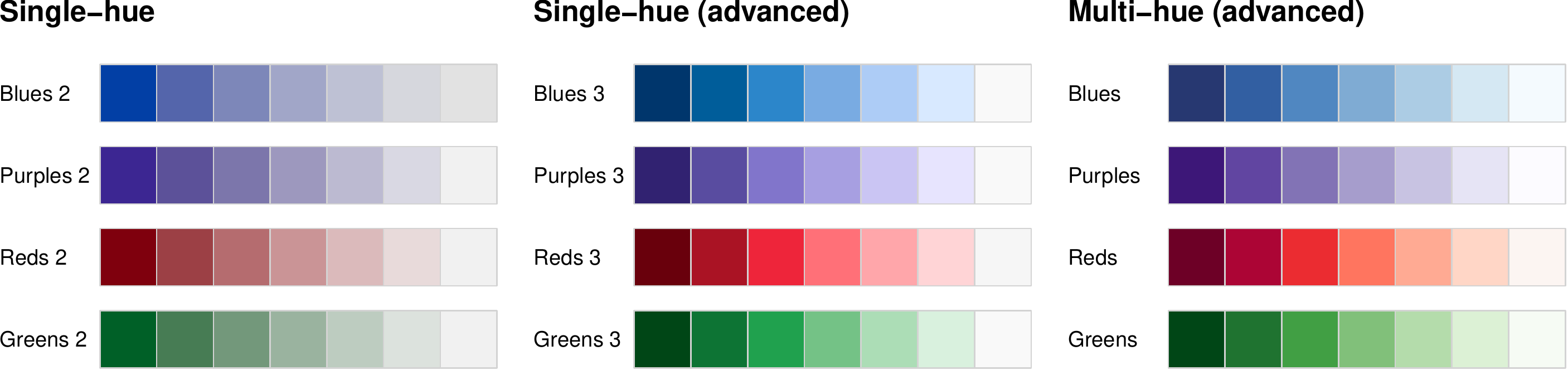} 
\caption[Variations of blue, purple, red, and green palettes with single hue and monotonic chroma (left), single hue and triangular chroma (center), and multiple hues and triangular chroma (right)]{Variations of blue, purple, red, and green palettes with single hue and monotonic chroma (left), single hue and triangular chroma (center), and multiple hues and triangular chroma (right).}\label{fig:swatch-brpg}
\end{figure}

Next, we demonstrate a more complex example of a \texttt{swatchplot()}
with three matrices of sequential color palettes of blues, purples,
reds, and greens (see Figure~\ref{fig:swatch-brpg}). For all palettes,
luminance increases monotonically to yield a proper sequential palette.
However, the hue and chroma handling is somewhat different to emphasize
different parts of the palette.
\begin{itemize}
\tightlist
\item
  \emph{Single-hue:} In each palette the hue is fixed and chroma
  decreases monotonically (along with increasing luminance). This is
  typically sufficient to clearly bring out the extreme colors
  (dark/colorful vs.~light gray).
\item
  \emph{Single-hue (advanced):} The hue is fixed (as above) but the
  chroma trajectory is triangular. Compared to the basic single-hue
  palette above this allows to better distinguish the colors in the
  middle and not only the extremes.
\item
  \emph{Multi-hue (advanced):} As in the advanced single-hue palette the
  chroma trajectory is triangular but additionally the hue varies
  slightly. This can further enhance the distinction of colors in the
  middle of the palette.
\end{itemize}
\begin{CodeChunk}
\begin{CodeInput}
R> bprg <- c("Blues", "Purples", "Reds", "Greens")
R> swatchplot(
+  "Single-hue"     = t(sapply(paste(bprg, 2), sequential_hcl, n = 7)),
+  "Single-hue (advanced)" = t(sapply(paste(bprg, 3), sequential_hcl, n = 7)),
+  "Multi-hue (advanced)"  = t(sapply(bprg,    sequential_hcl, n = 7)),
+  nrow = 5, line = 5)
\end{CodeInput}
\end{CodeChunk}

\subsection{HCL (and RGB) spectrum}\label{sec:specplot}

As the properties of a palette in terms of the perceptual dimensions
\emph{hue}, \emph{chroma}, and \emph{luminance} are not always clear
from looking just at color swatches or (statistical) graphics based on
these palettes, the \texttt{specplot()} function provides an explicit
display for the coordinates of the HCL trajectory associated with a
palette. This can bring out clearly various aspects, e.g., whether hue
is constant, chroma is monotonic or triangular, and whether luminance is
approximately constant (as in many qualitative palettes), monotonic (as
in sequential palettes), or diverging.

The function first transforms a given color palette to its HCL
(\texttt{polarLUV()}) coordinates. As the hues for low-chroma colors are
not (or only poorly) identified, by default a smoothing is applied to
the hues. Also, to avoid jumps from 0 to 360 or vice versa, the hue
coordinates are shifted suitably. By default, the resulting trajectories
in the HCL spectrum are visualized by a simple line plot:
\begin{itemize}
\tightlist
\item
  Hue is drawn in red and coordinates are indicated on the axis on the
  right with range \([0, 360]\) or (if necessary) \([-360, 360]\).
\item
  Chroma is drawn in green with coordinates on the left axis. The range
  {[}0, 100{]} is used unless the palette necessitates higher chroma
  values.
\item
  Luminance is drawn in blue with coordinates on the left axis in the
  range {[}0, 100{]}.
\end{itemize}
Additionally, a color swatch for the palette is included. Optionally, a
second spectrum for the corresponding trajectories of RGB coordinates
can be included. However, this is usually just of interest for palettes
created in RGB space (or simple transformations of RGB).

\begin{figure}[t!]
\centering
\includegraphics[width=0.49\textwidth]{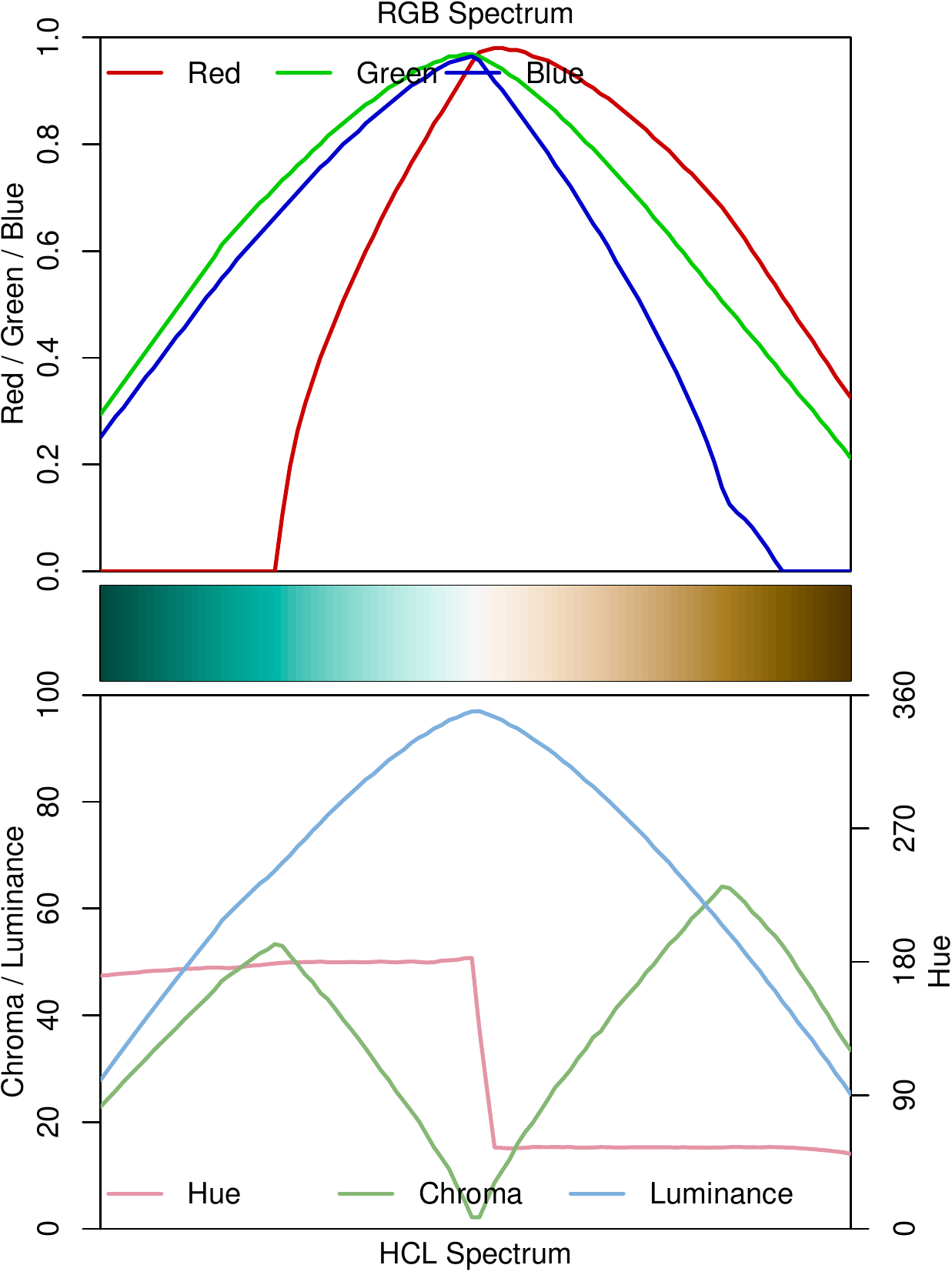} \includegraphics[width=0.49\textwidth]{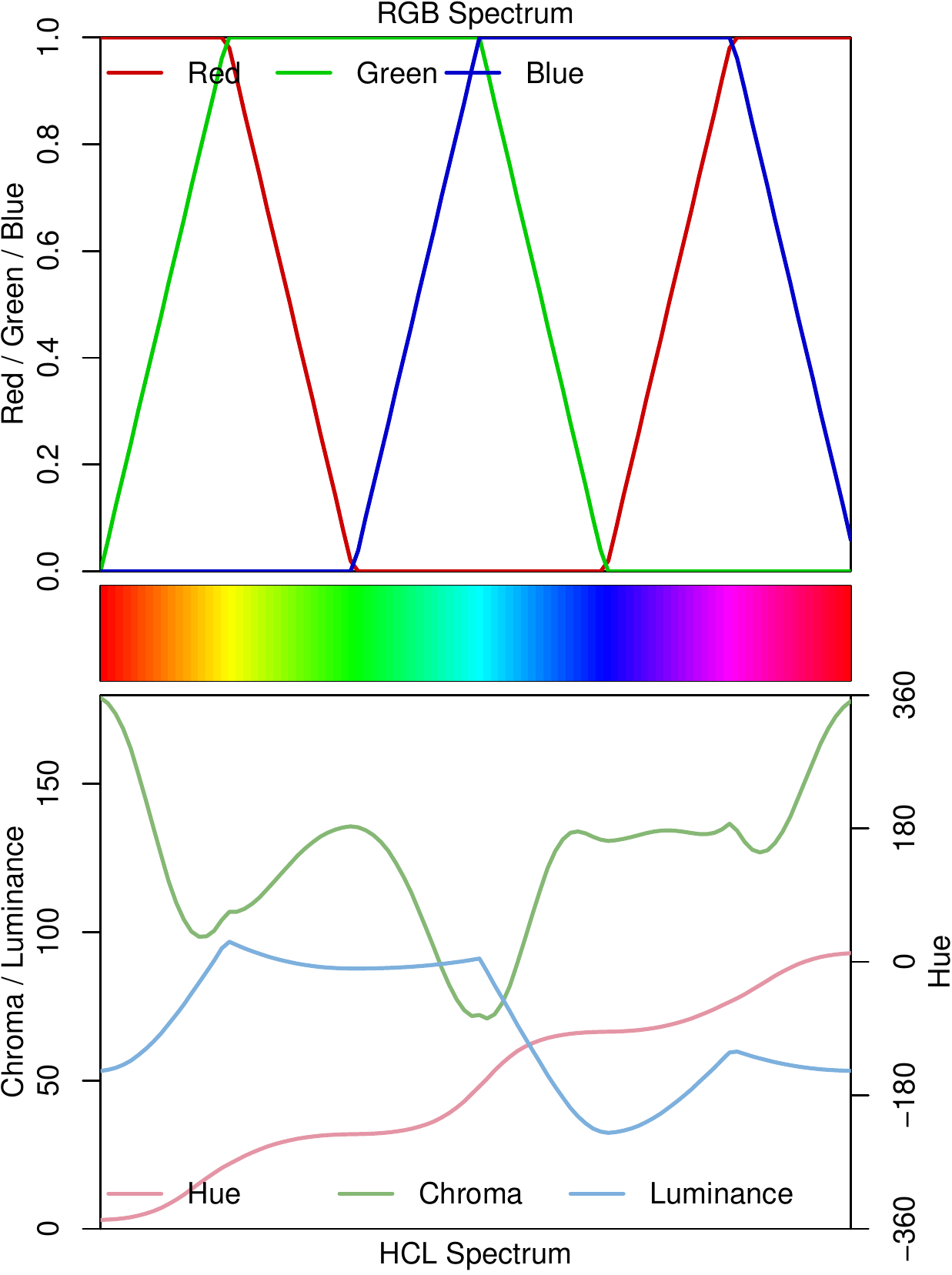} 
\caption[HCL spectrum of the balanced diverging \code{"Green-Brown"} palette (left panel) and the (in)famous and rather unbalanced \code{rainbow()} palette (right panel)]{HCL spectrum of the balanced diverging \code{"Green-Brown"} palette (left panel) and the (in)famous and rather unbalanced \code{rainbow()} palette (right panel).}\label{fig:specplot}
\end{figure}

As spectrum plots have already been used for illustration in
Figures~\ref{fig:allplots-qualitative} (for a qualitative palette) as
well as \ref{fig:allplots-sequential} and \ref{fig:brewer-carto-viridis}
(for sequential palettes), this section only provides a couple of
additional illustrations. The diverging \texttt{"Green-Brown"} palette
is depicted in the left panel of Figure~\ref{fig:specplot}. It simply
combines a green and a brown/yellow sequential single-hue palette, both
with triangular chroma trajectory. Hue is constant in each ``arm'' of
the palette and the chroma/luminance trajectories are rather balanced
between both arms. In the center the palette passes through a light gray
(with zero chroma) as the neutral value. By including the corresponding
RGB spectrum in the top panel, it also becomes apparent that choosing
such well-balanced palettes through trajectories in RGB color space is
not straightforward. This balanced palette -- based on relatively simple
HCL trajectories -- is contrasted with a poorly-balanced palette --
based on simple linear RGB trajectories in the right panel of
Figure~\ref{fig:specplot}. This depicts the RGB and HCL spectrum of the
(in)famous RGB rainbow palette. \citep[See][ for a plea why the RGB
rainbow palette should be avoided in almost all scientific
graphics.]{color:Hawkins+McNeall+Stephenson:2014}
\begin{CodeChunk}
\begin{CodeInput}
R> specplot(diverging_hcl(100, "Green-Brown"), rgb = TRUE)
R> specplot(rainbow(100), rgb = TRUE)
\end{CodeInput}
\end{CodeChunk}
The RGB spectrum of the rainbow palette shows that the trajectories are
quite simple in RGB space but lead to substantial variations in chroma
and (more importantly) luminance. This is why this palette is not
suitable for encoding underlying data in statistical graphics. See also
the related discussion of color vision deficiency in
Section~\ref{sec:color_vision_deficiency}.

\subsection{Trajectories in HCL space}\label{sec:hclplot}

\begin{figure}[p!]
\centering
\includegraphics[width=0.4\textwidth]{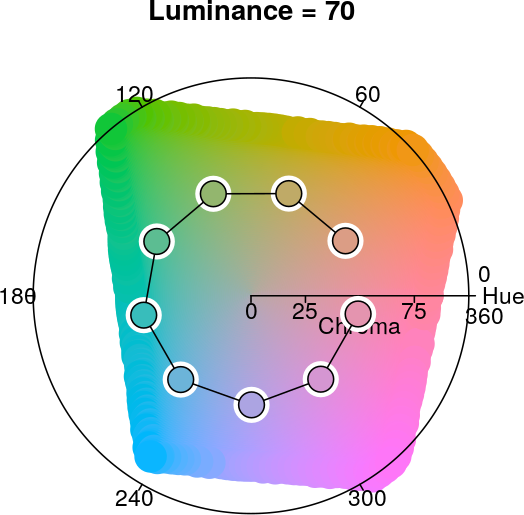} 
\caption[Hue-chroma plane with luminance fixed at $L = 70$ along with the qualitative \texttt{"Dynamic"} palette with varying hue $H$ and chroma fixed at $C = 50$]{Hue-chroma plane with luminance fixed at $L = 70$ along with the qualitative \texttt{"Dynamic"} palette with varying hue $H$ and chroma fixed at $C = 50$.}\label{fig:hcl-qualitative}

\vspace*{0.5cm}

\includegraphics[width=\textwidth]{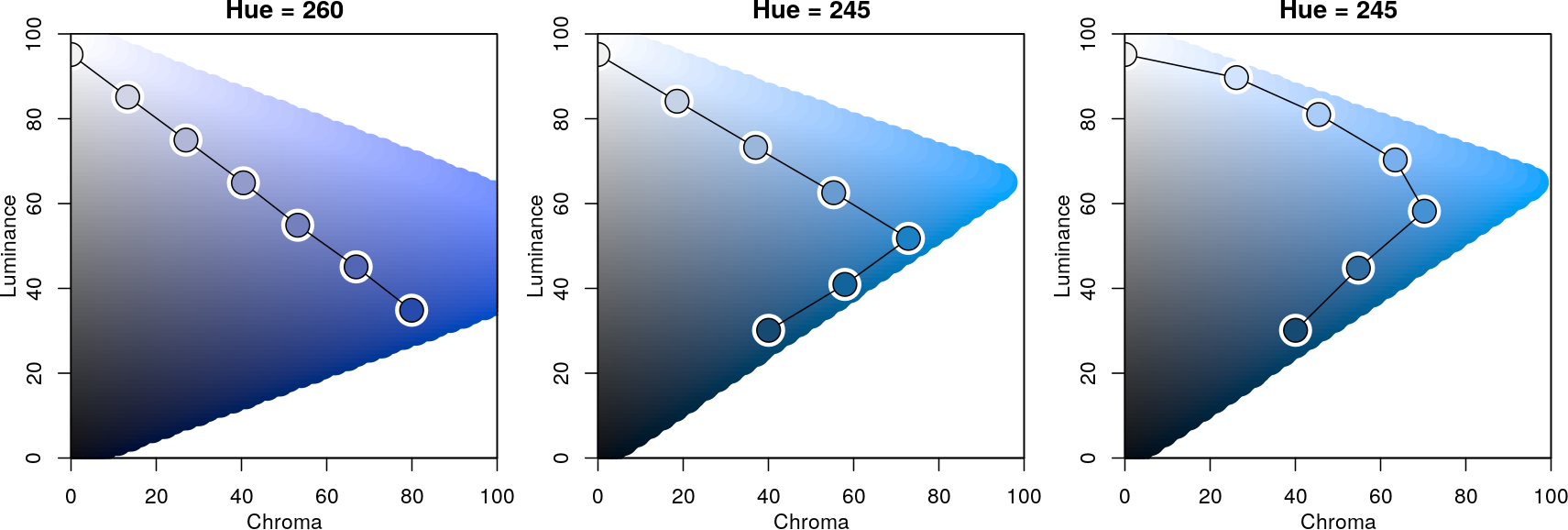} 
\caption[Luminance-chroma plane with variations of blue sequential single-hue palettes (similar to \texttt{"Blues 2"} and \texttt{"Blues 3"})]{Luminance-chroma plane with variations of blue sequential single-hue palettes (similar to \texttt{"Blues 2"} and \texttt{"Blues 3"}). Left: Linear chroma for $H = 260$. Center: Triangular chroma for $H = 245$. Right: Power-transformed triangular chroma for $H = 245$.}\label{fig:hcl-sequential}

\vspace*{0.5cm}

\includegraphics[width=\textwidth]{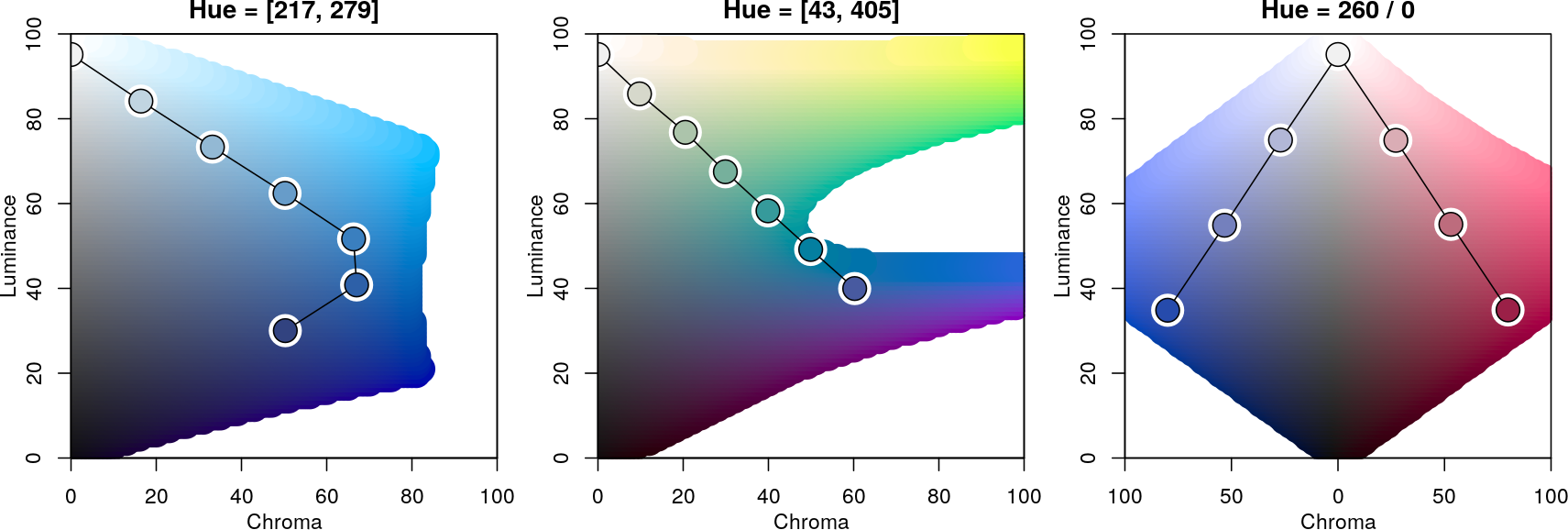} 
\caption[Luminance-chroma plane with blue multi-hue palette and triangular chroma (left), blue-yellow multi-hue palette and linear chroma (center), and diverging blue-red palette with balanced linear chroma]{Luminance-chroma plane with blue multi-hue palette and triangular chroma (left), blue-yellow multi-hue palette and linear chroma (center), and diverging blue-red palette with balanced linear chroma.}\label{fig:hcl-multi}
\end{figure}

While the \texttt{specplot()} function above works well for bringing out
the HCL coordinates associated with a given palette, it does not show
how the palette fits into the HCL space. For example, it is not so clear
whether high chroma values are close to the maximum possible for a given
hue. Thus, it cannot be judged so easily how the parameters of the hue,
chroma, and luminance trajectories can be modified to obtain another
palette.

Therefore, the \texttt{hclplot()} is another visualization of the HCL
coordinates associated with a palette. It does so by collapsing over one
of the coordinates (either the hue \(H\) or the luminance \(L\)) and
displays a heatmap of colors combining the remaining two dimensions. The
coordinates for the given color palette are highlighted to bring out its
trajectory. In case the hue is really fixed (as in single-hue sequential
palettes) or the luminance is really fixed (as in the qualitative
palettes), collapsing is straightforward. However, when the coordinate
that is collapsed over is actually not constant in the palette, a simple
bivariate linear model is used to capture how the collapsed coordinate
varies along with the two displayed coordinates.

The function \texttt{hclplot()} has been designed to work well with the
\texttt{hcl\_palettes()} in this package. While it is possible to apply
it to other color palettes as well, the results might look weird or
confusing if these palettes are constructed very differently (e.g., like
the highly-saturated base \proglang{R} palettes). To infer the default
\texttt{type} of projection \texttt{hclplot()} assesses the luminance
trajectory and sets the default correspondingly:
\begin{itemize}
\tightlist
\item
  \texttt{type\ =\ "qualitative"} if luminance is approximately
  constant.
\item
  \texttt{type\ =\ "sequential"} if luminance is monotonic.
\item
  \texttt{type\ =\ "diverging"} if luminance is diverging with two
  monotonic ``arms'' in the trajectory.
\end{itemize}
Thus, for qualitative palettes -- where luminance and chroma are fixed
-- the varying hue is displayed in a projection onto the hue-chroma
plane at a given fixed luminance (Figure~\ref{fig:hcl-qualitative}):
\begin{CodeChunk}
\begin{CodeInput}
R> hclplot(qualitative_hcl(9, "Dynamic"))
\end{CodeInput}
\end{CodeChunk}
Figure~\ref{fig:hcl-sequential} compares three single-hue sequential
palettes by projection to the luminance-chroma plane for the given fixed
hue. In the left panel the hue 260 is used with a simple linear chroma
trajectory. The other two panels employ a triangular chroma trajectory
for hue 245, either with a piecewise-linear (center) or
power-transformed (right) trajectory.
\begin{CodeChunk}
\begin{CodeInput}
R> par(mfrow = c(1, 3))
R> hclplot(sequential_hcl(7, h = 260, c = 80, l = c(35, 95), power = 1))
R> hclplot(sequential_hcl(7, h = 245, c = c(40, 75, 0), l = c(30, 95),
+    power = 1))
R> hclplot(sequential_hcl(7, h = 245, c = c(40, 75, 0), l = c(30, 95),
+    power = c(0.8, 1.4)))
\end{CodeInput}
\end{CodeChunk}
Note that for \(H = 260\) it is possible to go to dark colors (= low
luminance) with high chroma while this is not possible to the same
extent for \(H = 245\) due to the distorted shape of the~HCL space.
Hence, chroma has to be decreased when proceeding to the dark
low-luminance colors.

Finally, Figure~\ref{fig:hcl-multi} compares two multi-hue sequential
palettes along with a diverging palette.
\begin{CodeChunk}
\begin{CodeInput}
R> par(mfrow = c(1, 3))
R> hclplot(sequential_hcl(7, h = c(260, 220), c = c(50, 75, 0),
+    l = c(30, 95), power = 1))
R> hclplot(sequential_hcl(7, h = c(260, 60), c = 60, l = c(40, 95),
+    power = 1))
R> hclplot(diverging_hcl(7, h = c(260, 0), c = 80, l = c(35, 95),
+    power = 1))
\end{CodeInput}
\end{CodeChunk}
The multi-hue palette on the left employs a small hue range, resulting
in a palette of ``blues'' just with slightly more distinction of the
middle colors in the palette. In contrast, the multi-hue ``blue-yellow''
palette in the center panel uses a large hue range, resulting in more
color contrasts throughout the palette. Finally, the balanced diverging
palette in the right panel is constructed from two simple single-hue
sequential palettes (for hues 260/blue and 0/red) that are completely
balanced between the two ``arms'' of the palette.

\subsection{Demonstration of statistical graphics}\label{sec:demoplot}

\begin{figure}[t!]
\centering
\includegraphics[width=\textwidth]{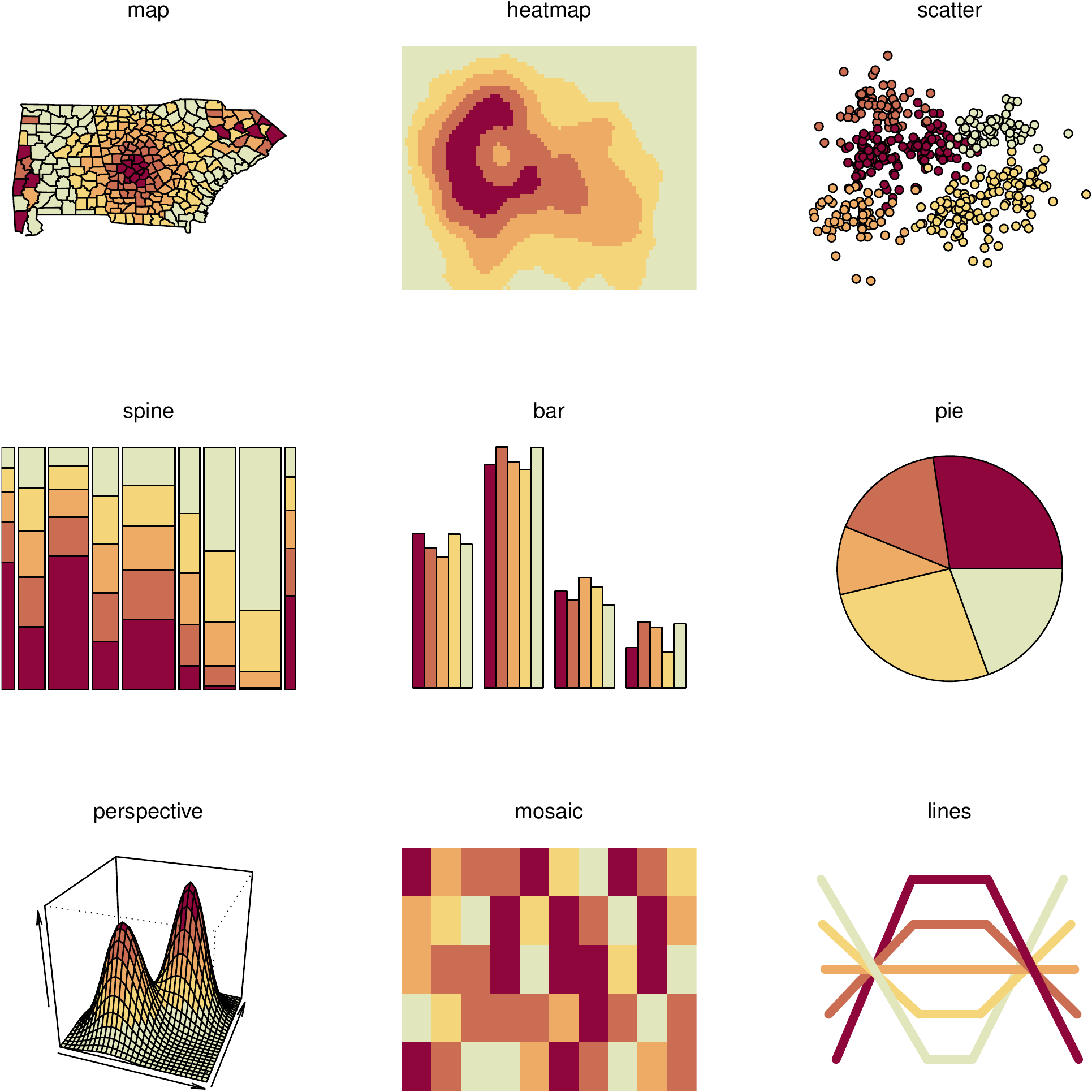} 
\caption[All built-in \code{demoplot} types with the same \code{sequential\_hcl(5, "Heat")} palette]{All built-in \code{demoplot} types with the same \code{sequential\_hcl(5, "Heat")} palette.}\label{fig:demoplot-all}
\end{figure}

To demonstrate how different kinds of color palettes work in different
kinds of statistical displays, \texttt{demoplot()} provides a simple
convenience interface to some base graphics with (mostly artificial)
data sets. As a first overview, Figure~\ref{fig:demoplot-all} displays
all built-in demos with the same sequential heat colors palette:
\texttt{sequential\_hcl(5,\ "Heat")}. All types of demos can, in
principle, deal with arbitrarily many colors from any palette but
clearly the graphics differ in various respects such as:
\begin{itemize}
\tightlist
\item
  Work best for fewer colors (e.g., bar, pie, scatter, lines, \ldots{})
  vs.~many colors (e.g., heatmap, perspective, \ldots{}).
\item
  Intended for categorical data (e.g., bar, pie, \ldots{})
  vs.~continuous numeric data (e.g., heatmap, perspective, \ldots{}).
\item
  Shading areas (e.g., map, bar, pie, \ldots{}) vs.~shading points or
  lines (scatter, lines).
\end{itemize}
Hence, in the following some further illustrations are organized by type
of palette, using suitable demos for the particular palettes.

\emph{Qualitative palettes:} Light pastel colors typically work better
for shading areas (pie, left) while darker and more colorful palettes
are usually preferred for points (center) or lines (right).
\begin{CodeChunk}
\begin{CodeInput}
R> par(mfrow = c(1, 3))
R> demoplot(qualitative_hcl(4, "Pastel 1"), type = "pie")
R> demoplot(qualitative_hcl(4, "Set 2"),    type = "scatter")
R> demoplot(qualitative_hcl(4, "Dark 3"),   type = "lines")
\end{CodeInput}
\end{CodeChunk}
\begin{figure}[t!]
\centering
\includegraphics[width=\textwidth]{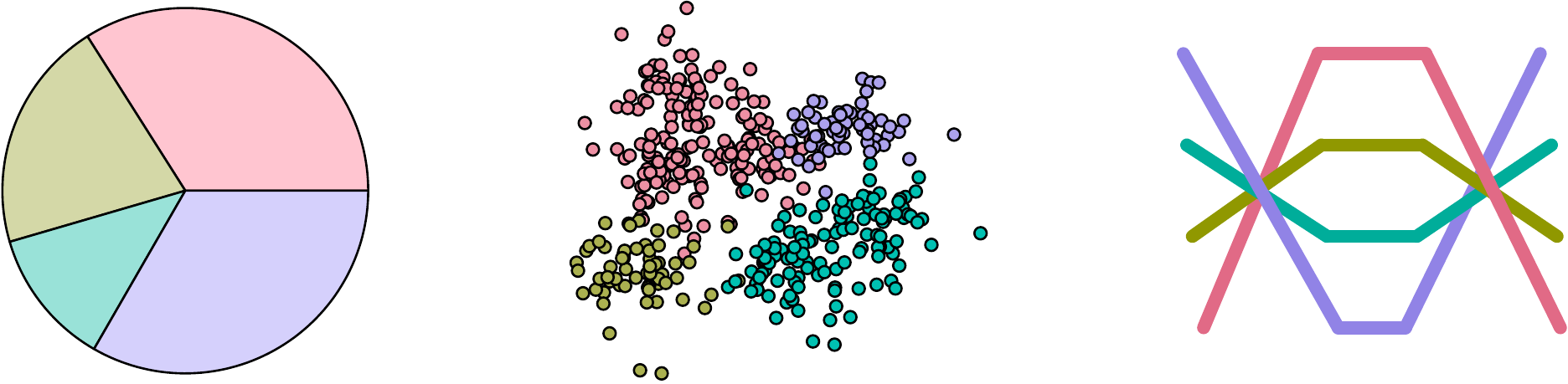} 
\caption[Examples for \code{demoplot()} with different \code{qualitative\_hcl()} palettes]{Examples for \code{demoplot()} with different \code{qualitative\_hcl()} palettes.}\label{fig:demoplot-qualitative}
\end{figure}
\begin{figure}[t!]
\centering
\includegraphics[width=\textwidth]{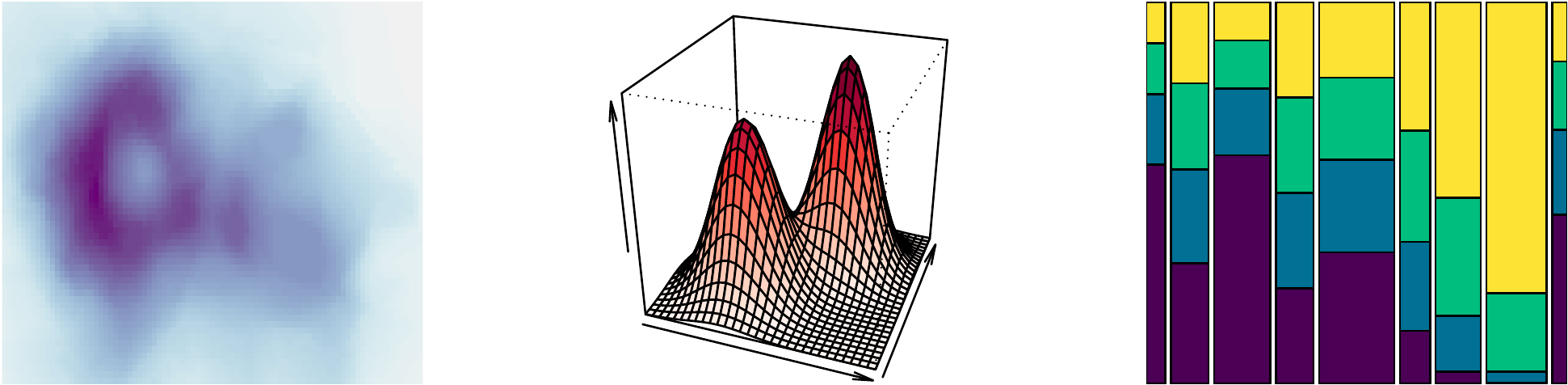} 
\caption[Examples for \code{demoplot()} with different \code{sequential\_hcl()} palettes]{Examples for \code{demoplot()} with different \code{sequential\_hcl()} palettes.}\label{fig:demoplot-sequential}
\end{figure}
\begin{figure}[t!]
\centering
\includegraphics[width=\textwidth]{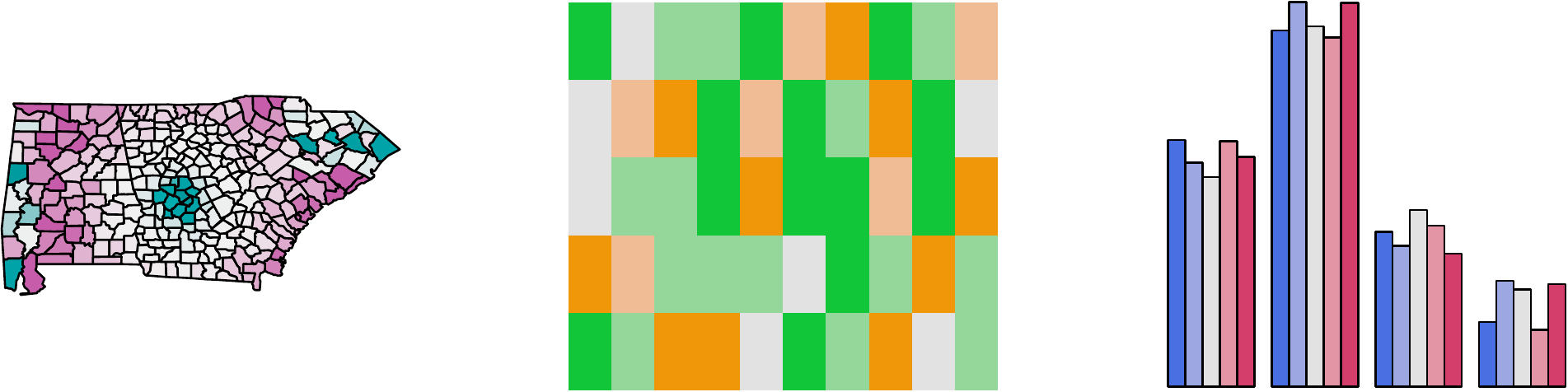} 
\caption[Examples for \code{demoplot()} with different \code{diverging\_hcl()} palettes]{Examples for \code{demoplot()} with different \code{diverging\_hcl()} palettes.}\label{fig:demoplot-diverging}
\end{figure}
\emph{Sequential palettes:} Heatmaps (left) or perspective plots
(center) often employ almost continuous gradients with strong luminance
contrasts. In contrast, when only a few ordered categories are to be
displayed (e.g., in a spine plot, right) more colorful sequential
palettes like the viridis palette can be useful.
\begin{CodeChunk}
\begin{CodeInput}
R> par(mfrow = c(1, 3))
R> demoplot(sequential_hcl(99, "Purple-Blue"), type = "heatmap")
R> demoplot(sequential_hcl(99, "Reds"),        type = "perspective")
R> demoplot(sequential_hcl( 4, "Viridis"),     type = "spine")
\end{CodeInput}
\end{CodeChunk}
\emph{Diverging palettes:} In some displays (such as the map, left), it
is useful to employ an almost continuous gradient with strong luminance
contrast to bring out the extremes. Here, this contrast is amplified by
a larger power transformation emphasizing the extremes even further. In
contrast, when fewer colors are needed more colorful palettes with lower
luminance contrasts can be desired. This is exemplified by a mosaic
(center) and bar plot (right).
\begin{CodeChunk}
\begin{CodeInput}
R> par(mfrow = c(1, 3))
R> demoplot(diverging_hcl(99, "Tropic", power = 2.5), type = "map")
R> demoplot(diverging_hcl( 5, "Green-Orange"),        type = "mosaic")
R> demoplot(diverging_hcl( 5, "Blue-Red 2"),          type = "bar")
\end{CodeInput}
\end{CodeChunk}
\begin{figure}[t!]
\centering
\includegraphics[width=\textwidth]{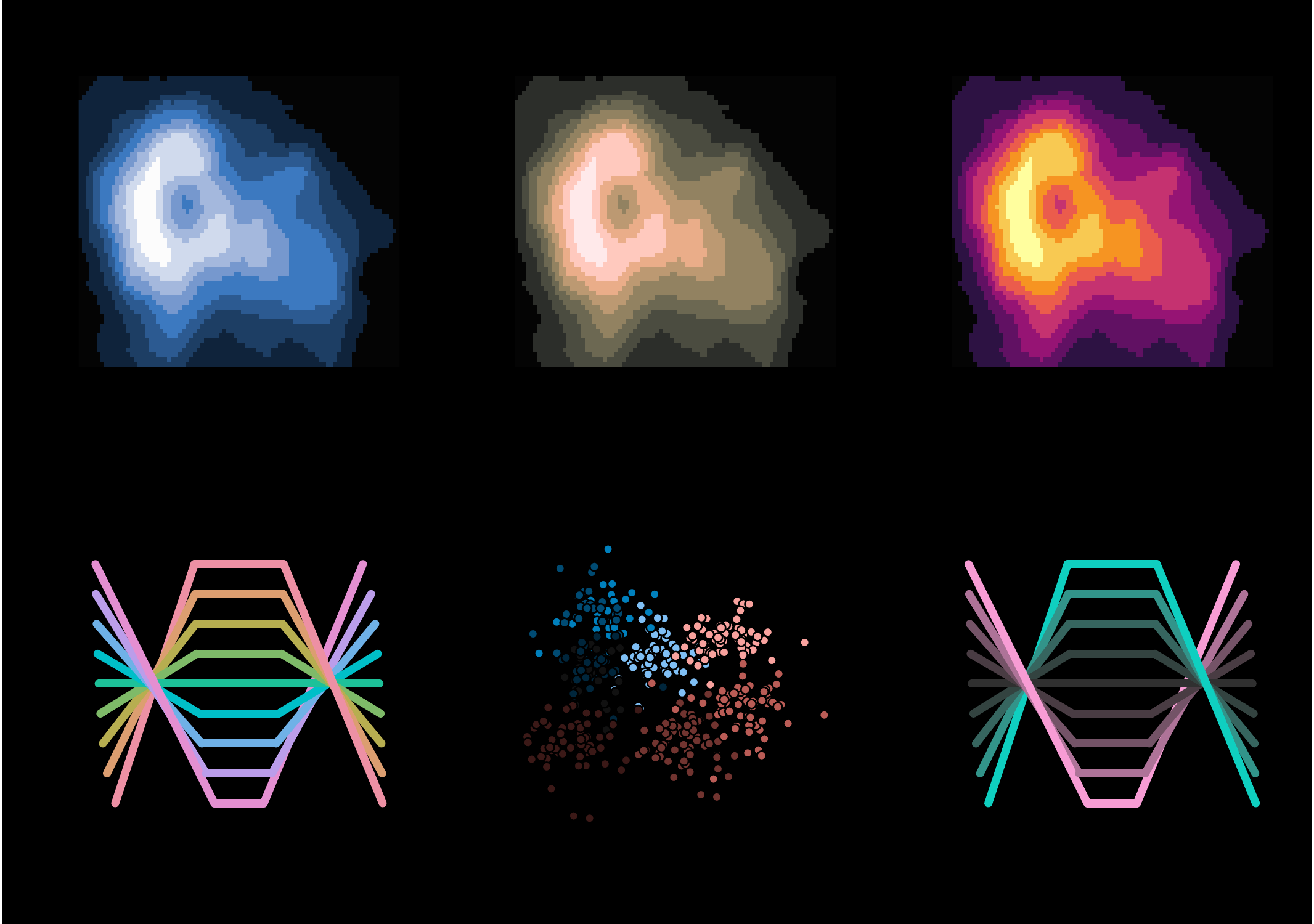} 
\caption[Examples for \code{demoplot()} with different palettes that work well on a black/dark background]{Examples for \code{demoplot()} with different palettes that work well on a black/dark background.}\label{fig:demoplot-dark}
\end{figure}
Figures~\ref{fig:demoplot-qualitative}--\ref{fig:demoplot-diverging}
focused on palettes designed for light/white backgrounds. Therefore, to
conclude, some palettes are highlighted in
Figure~\ref{fig:demoplot-dark} that work well on dark/black backgrounds.
\begin{CodeChunk}
\begin{CodeInput}
R> par(mfrow = c(2, 3), bg = "black")
R> demoplot(sequential_hcl(9, "Oslo"), "heatmap")
R> demoplot(sequential_hcl(9, "Turku"), "heatmap")
R> demoplot(sequential_hcl(9, "Inferno", rev = TRUE), "heatmap")
R> demoplot(qualitative_hcl(9, "Set 2"), "lines")
R> demoplot(diverging_hcl(9, "Berlin"), "scatter")
R> demoplot(diverging_hcl(9, "Cyan-Magenta", l2 = 20), "lines")
\end{CodeInput}
\end{CodeChunk}

\pagebreak

\section{Color vision deficiency emulation}\label{sec:color_vision_deficiency}

Using the physiologically-based model for simulating color vision
deficiency (CVD) of \citet{color:Machado+Oliveira+Fernandes:2009}
different kinds of limitations can be emulated: deuteranope (green cone
cells defective), protanope (red cone cells defective), and tritanope
(blue cone cells defective).

Below we briefly describe our \proglang{R} interface to these emulation
techniques and show them in practice for a heatmap with sequential
palette. Another diverging palette example is available at
\url{http://colorspace.R-Forge.R-project.org/articles/color_vision_deficiency.html}.
Finally, CVD emulation is particularly useful for bringing out why the
RGB rainbow palette is almost always a bad choice in scientific
displays. See
\url{http://colorspace.R-Forge.R-project.org/articles/endrainbow.html}
for further illustrations.

\subsection[R functions]{\proglang{R} functions}\label{r-functions}

The workhorse function to emulate color-vision deficiency is
\texttt{simulate\_cvd()}, which can take any vector of valid
\proglang{R} colors and transform them according to a certain CVD
transformation matrix and transformation equation. The transformation
matrices have been established by
\citet{color:Machado+Oliveira+Fernandes:2009} and are provided in
objects \texttt{protanomaly\_cvd}, \texttt{deutanomaly\_cvd}, and
\texttt{tritanomaly\_cvd}. The convenience interfaces \texttt{deutan()},
\texttt{protan()}, and \texttt{tritan()} are the high-level functions
for simulating the corresponding kind of color blindness with a given
severity (calling \texttt{simulate\_cvd()} internally).

For further guidance on color blindness in relation to statistical
graphics see \citet{color:Lumley:2006} which accompanies the
\proglang{R} package \pkg{dichromat} \citep{color:dichromat} and is
based on earlier emulation techniques
\citep{color:Vienot+Brettel+Ott:1995, color:Brettel+Vienot+Mollon:1997, color:Vienot+Brettel+Mollon:1999}.

\subsection{Illustration: Heatmap with sequential palette}\label{illustration-heatmap-with-sequential-palette}

\begin{figure}[b!]
\centering
\includegraphics[width=0.5\textwidth]{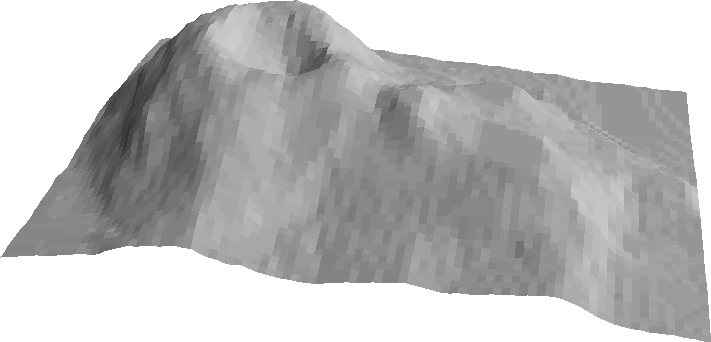} 
\caption[Perspective visualization of Maunga Whau \code{volcano} data (Mount Eden, Auckland, New Zealand)]{Perspective visualization of Maunga Whau \code{volcano} data (Mount Eden, Auckland, New Zealand).}\label{fig:volcano}
\end{figure}

\begin{figure}[p!]
\centering
\includegraphics[width=0.8\textwidth]{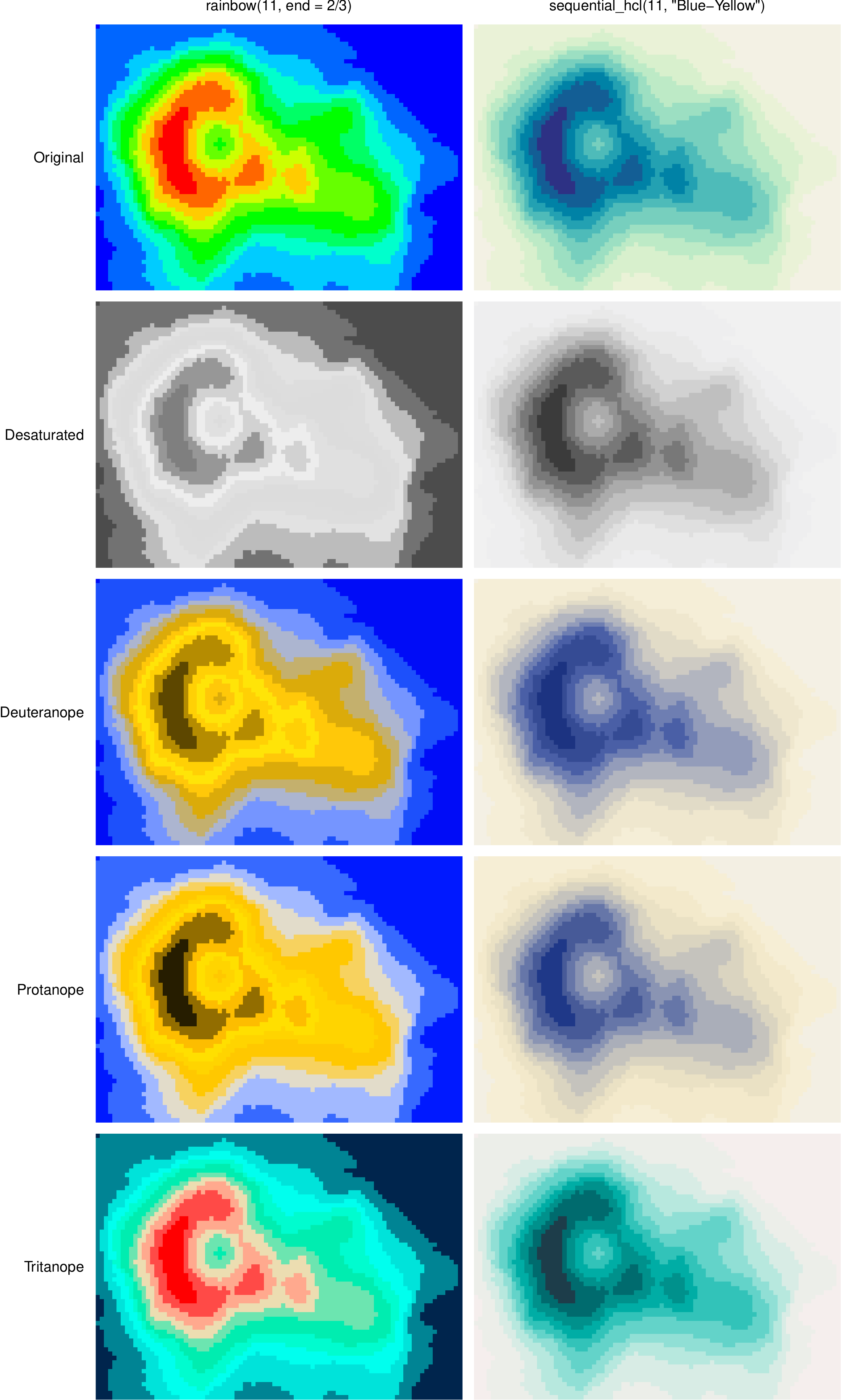} 
\caption[Heatmap of Maunga Whau \code{volcano} data with RGB rainbow (left) and HCL-based blue-yellow palette (right)]{Heatmap of Maunga Whau \code{volcano} data with RGB rainbow (left) and HCL-based blue-yellow palette (right). The first row shows the original color palettes while subsequent rows emulate various color deficiencies.}\label{fig:heatmap-sequential}
\end{figure}

To illustrate that poor color choices can severely reduce the usefulness
of a statistical graphic for readers with color vision deficiencies, we
employ the infamous RGB rainbow color palette in a heatmap. In base
\proglang{R} this can be generated by \texttt{rainbow(11,\ end\ =\ 2/3)}
ranging from red (for high values) to blue (for low values). The poor
results for the RGB rainbow palette are contrasted in
Figure~\ref{fig:heatmap-sequential} with a proper sequential palette
ranging from dark blue to light yellow:
\texttt{sequential\_hcl(11,\ "Blue-Yellow")}.

The statistical graphic employed for illustration is a heatmap of the
well-known Maunga Whau \texttt{volcano} data from base \proglang{R}.
This heatmap is easily available as \texttt{demoplot(...,\ "heatmap")}
where \texttt{...} is the color vector to be used, e.g.,
\begin{CodeChunk}
\begin{CodeInput}
R> rainbow(11, end = 2/3)
\end{CodeInput}
\begin{CodeOutput}
 [1] "#FF0000FF" "#FF6600FF" "#FFCC00FF" "#CCFF00FF" "#66FF00FF"
 [6] "#00FF00FF" "#00FF66FF" "#00FFCCFF" "#00CCFFFF" "#0066FFFF"
[11] "#0000FFFF"
\end{CodeOutput}
\begin{CodeInput}
R> deutan(rainbow(11, end = 2/3))
\end{CodeInput}
\begin{CodeOutput}
 [1] "#5D4700FF" "#B58C01FF" "#FFD005FF" "#FFE408FF" "#FFC809FF"
 [6] "#DBAB0AFF" "#C4B06DFF" "#ACB5D0FF" "#7595FFFF" "#1D50FBFF"
[11] "#000CF7FF"
\end{CodeOutput}
\end{CodeChunk}
and so on. To aid the interpretation of the heatmap a perspective
display using only gray shades is provided in Figure~\ref{fig:volcano},
providing another intuitive display of what the terrain around Maunga
Whau looks like.

Subsequently, all combinations of palette and color vision deficiency
are visualized. Additionally, a grayscale version is created with
\texttt{desaturate()}. This clearly shows how poorly the RGB rainbow
performs, often giving quite misleading impressions of what the terrain
around Maunga Whau looks like. In contrast, the HCL-based blue-yellow
palette works reasonably well in all settings. The most important
problem of the RGB rainbow is that it is not monotonic in luminance,
making correct interpretation quite hard. Moreover, the red-green
contrasts deteriorate substantially in the dichromatic emulations.

\section{Apps for choosing colors and palettes interactively}\label{sec:hclwizard}

To facilitate exploring the package and employing it when working with
colors, several graphical user interfaces (GUIs) are provided within the
package as \pkg{shiny} apps \citep{color:shiny}. All of these GUIs/apps
can either be run locally from within \proglang{R} but are also provided
online at \url{http://hclwizard.org/}.
\begin{itemize}
\tightlist
\item
  \emph{Palette constructor:} \texttt{choose\_palette()} or
  \texttt{hclwizard()} or \texttt{hcl\_wizard()}.
\item
  \emph{Color picker:} \texttt{choose\_color()} or equivalently
  \texttt{hcl\_color\_picker()}.
\item
  \emph{Color vision deficiency emulator:} \texttt{cvd\_emulator()}.
\end{itemize}
In addition to the \pkg{shiny} version, the \emph{palette constructor}
app is also available as a \proglang{Tcl}/\proglang{Tk} GUI via
\proglang{R} package \pkg{tcltk} shipped with base \proglang{R}
\citep{color:R}. The \pkg{tcltk} version can only be run locally and is
considerably faster while the \pkg{shiny} version has a nicer interface
with more features and can be run online. The \texttt{choose\_palette()}
function by default starts the \pkg{tcltk} version while
\texttt{hclwizard()}/\texttt{hcl\_wizard()} by default start the
\pkg{shiny} version.

\pagebreak

\subsection{Choose palettes with the HCL color model}\label{choose-palettes-with-the-hcl-color-model}

\begin{figure}[t!]
\centering
\includegraphics[width=\textwidth]{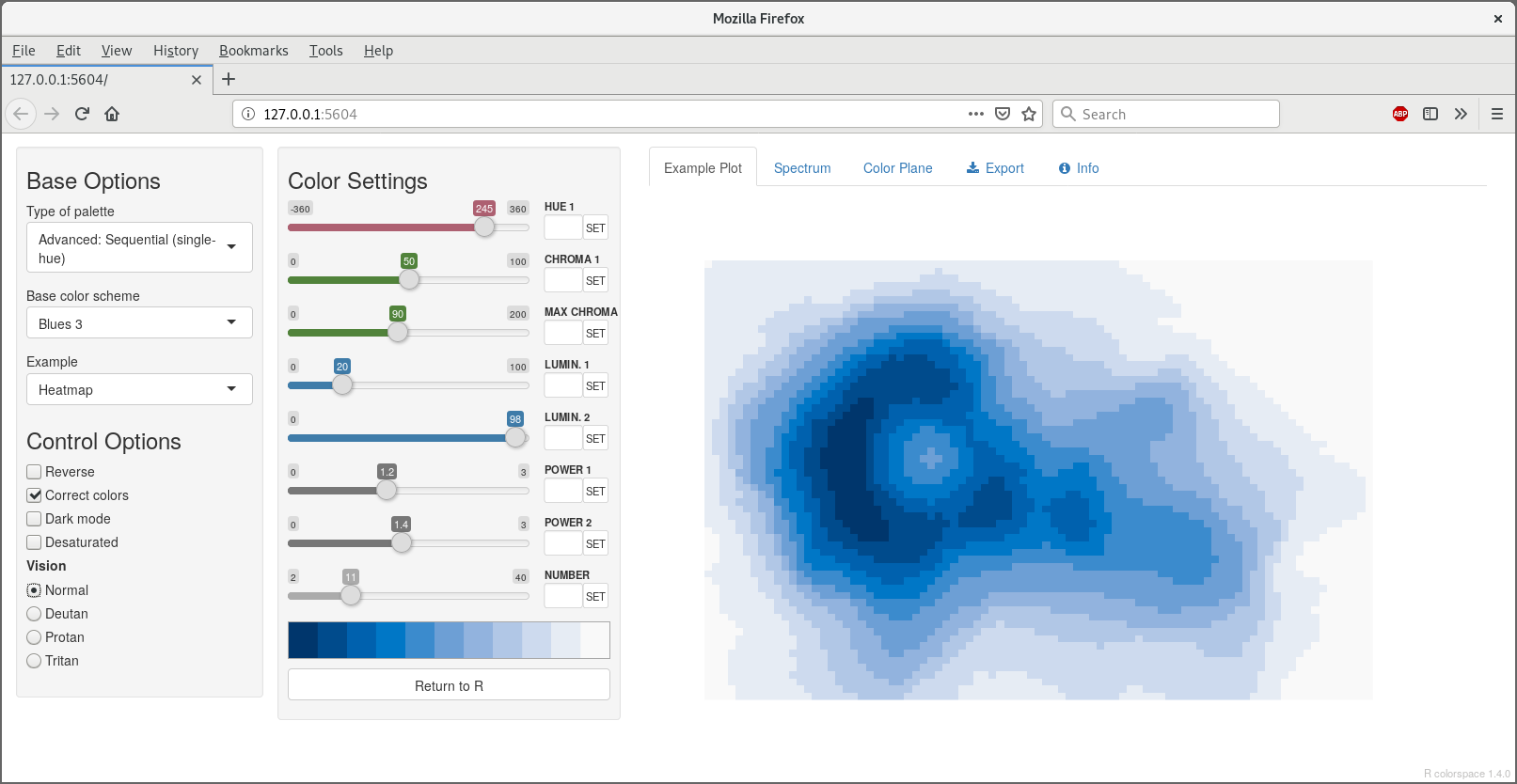} 
\caption[App for interactively choosing HCL-based color palettes]{App for interactively choosing HCL-based color palettes: \code{choose\_color()}/\code{hclwizard()}.}\label{fig:choose_palette}
\end{figure}

The \pkg{shiny} version of this GUI is shown in
Figure~\ref{fig:choose_palette}. It interfaces the
\texttt{qualitative\_hcl()}, \texttt{sequential\_hcl()}, and
\texttt{diverging\_hcl()} palettes from Section~\ref{sec:hcl_palettes}.
The GUIs allow for interactive modification of the arguments of the
respective palette-generating functions, i.e., starting/ending hue,
minimal/maximal chroma, minimal maximal luminance, and power
transformations that control how quickly/slowly chroma and/or luminance
are changed through the palette. Subsets of the parameters may not be
applicable depending on the type of palette chosen.

Optionally, the active palette can be illustrated by using a
\texttt{specplot()} (see Section~\ref{sec:specplot}), \texttt{hclplot()}
(see Section~\ref{sec:hclplot}), or \texttt{demoplot()} (see
Section~\ref{sec:demoplot}), and assessed using emulation of color
vision deficiencies (see Section~\ref{sec:color_vision_deficiency}). To
facilitate generation of palettes for black/dark backgrounds, a ``dark
mode'' of the GUIs is also available.

\subsection{Choose individual colors with the HCL color model}\label{choose-individual-colors-with-the-hcl-color-model}

\begin{figure}[t!]
\centering
\includegraphics[width=\textwidth]{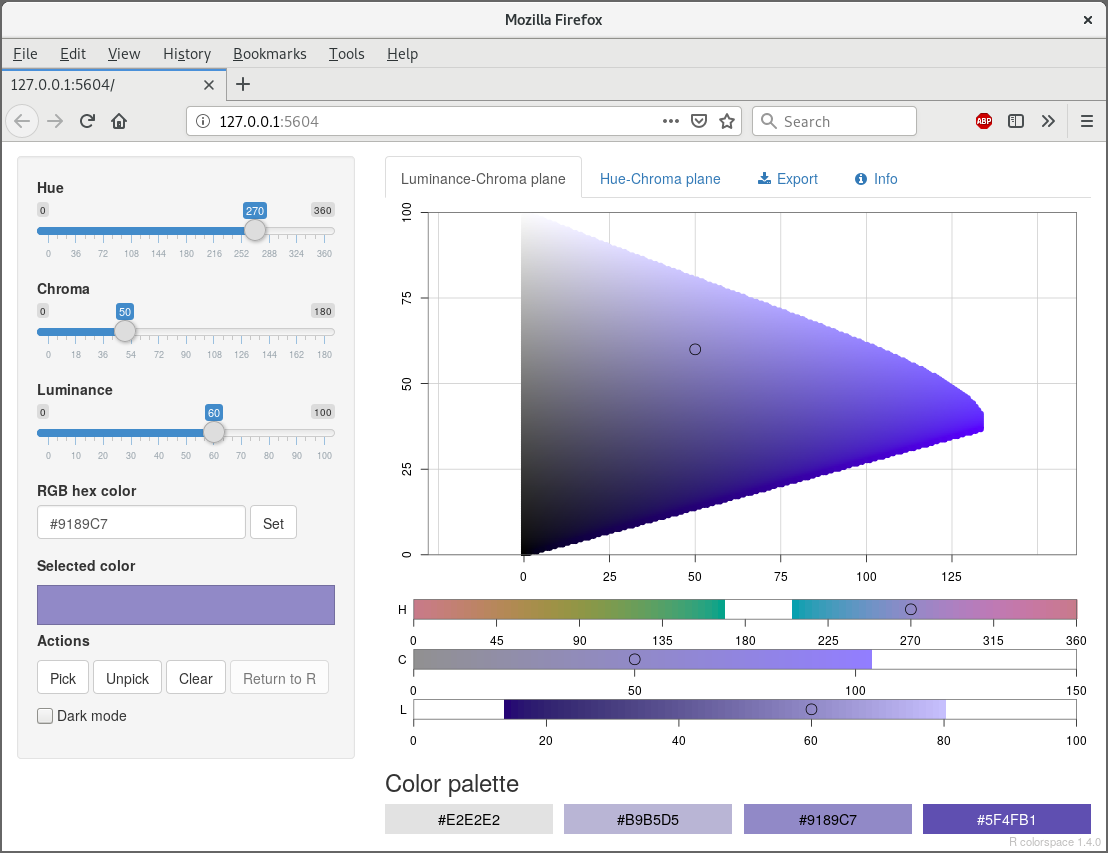} 
\caption[App for interactively choosing individual colors in HCL space]{App for interactively choosing individual colors in HCL space: \code{choose\_color()}/\code{hcl\_color\_picker()}.}\label{fig:choose_color}
\end{figure}

This GUI can be started with either \texttt{choose\_color()} or
equivalently \texttt{hcl\_color\_picker()}. It shows the HCL color space
either as a hue-chroma plane for a given luminance value or as a
luminance-chroma plane for a given hue. Colors can be entered by:
\begin{itemize}
\tightlist
\item
  Clicking on a color coordinate in the hue-chroma or luminance-chroma
  plane.
\item
  Specifying the hue/chroma/luminance values via sliders.
\item
  Entering an RGB hex code.
\end{itemize}
By repeating the selection a palette of colors can be constructed and
returned within \proglang{R} for subsequent usage in visualizations.

\subsection{Emulate color vision deficiencies}\label{emulate-color-vision-deficiencies}

\begin{figure}[t!]
\centering 
\includegraphics[width=0.49\textwidth]{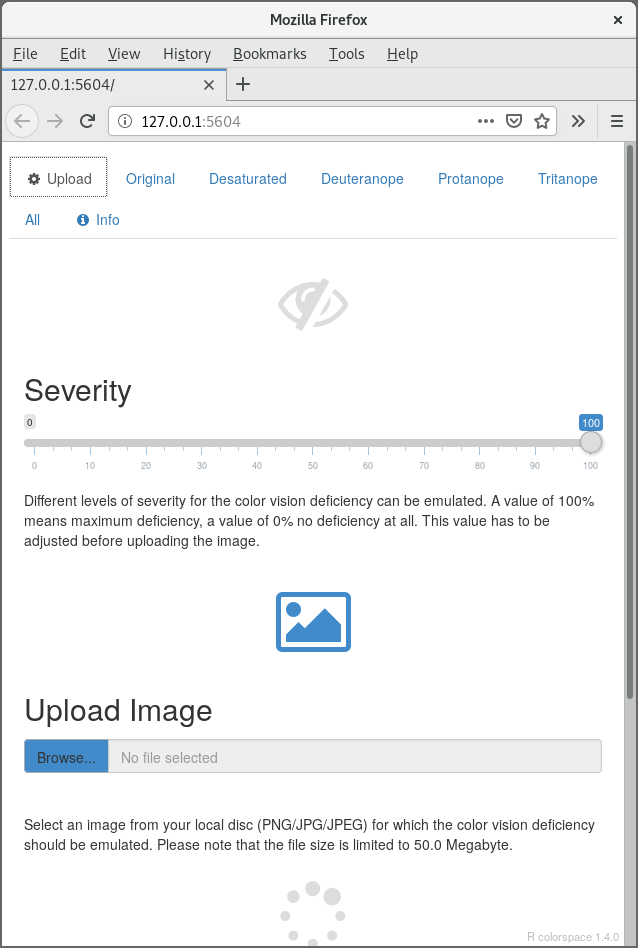} 
\includegraphics[width=0.49\textwidth]{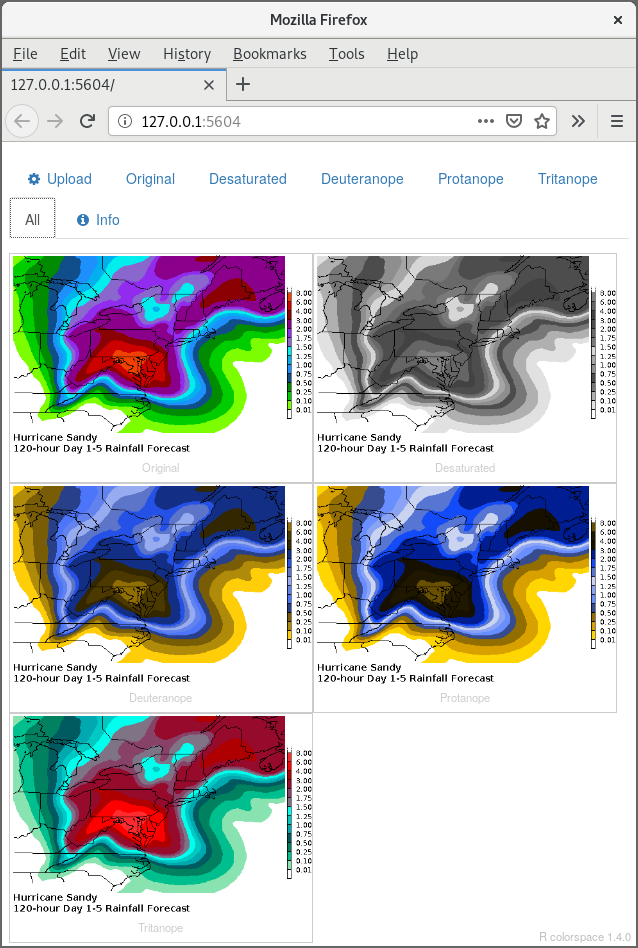}
\caption{App for emulating color vision deficiencies for uploaded raster images: \code{cvd\_emulator()}.}\label{fig:cvd_emulator}
\end{figure}

This GUI can be started with \texttt{cvd\_emulator()}. It allows to
upload a raster image in JPG or PNG format which is then checked for
various kinds of color vision deficiencies at the selected severity. By
default the severity is set to 100\% and all supported kinds of color
vision deficiency are checked for.

\section{Color manipulation and utilities}\label{sec:manipulation_utilities}

The \pkg{colorspace} package provides several color manipulation
utilities that are useful for creating, assessing, or transforming color
palettes, namely:
\begin{itemize}
\tightlist
\item
  \texttt{desaturate()}: Desaturate colors by chroma removal in HCL
  space.
\item
  \texttt{darken()} and \texttt{lighten()}: Algorithmically lighten or
  darken colors in HCL and/or HLS space.
\item
  \texttt{max\_chroma()}: Compute maximum chroma for given hue and
  luminance in HCL space.
\item
  \texttt{mixcolor()}: Additively mix two colors by computing their
  convex combination.
\end{itemize}

\subsection{Desaturation in HCL space}\label{desaturation-in-hcl-space}

Desaturation should map a given color to the gray with the same
``brightness''. In principle, any perceptually-based color model (HCL,
HLS, HSV, \ldots{}) could be employed for this but HCL works
particularly well because its coordinates capture the perceptual
properties better than most other color models.

The \texttt{desaturate()} function converts any given hex color code or
named \proglang{R} color to the corresponding HCL coordinates and sets
the chroma to zero. Thus, only the luminance matters which captures the
``brightness'' mentioned above. Finally, the resulting HCL coordinates
are transformed back to hex color codes for use in \proglang{R}. First,
\texttt{desaturate()} is used to desaturate a vector of \proglang{R}
color names:
\begin{CodeChunk}
\begin{CodeInput}
R> desaturate(c("white", "orange", "blue", "black"))
\end{CodeInput}
\begin{CodeOutput}
[1] "#FFFFFF" "#B8B8B8" "#4C4C4C" "#000000"
\end{CodeOutput}
\end{CodeChunk}
Notice that the hex codes corresponding to three coordinates in sRGB
space are always the same, indicating gray colors. Analogously, hex
color codes can also be transformed -- in this case RGB rainbow colors
from the base \proglang{R} function \texttt{rainbow()}:
\begin{CodeChunk}
\begin{CodeInput}
R> rainbow(3)
\end{CodeInput}
\begin{CodeOutput}
[1] "#FF0000FF" "#00FF00FF" "#0000FFFF"
\end{CodeOutput}
\begin{CodeInput}
R> desaturate(rainbow(3))
\end{CodeInput}
\begin{CodeOutput}
[1] "#7F7F7FFF" "#DCDCDCFF" "#4C4C4CFF"
\end{CodeOutput}
\end{CodeChunk}
Already this simple example shows that the three RGB rainbow colors have
very different grayscale levels. This can be brought even more clearly
when using a full color wheel (of colors with hues in {[}0, 360{]}
degrees). While the RGB \texttt{rainbow()} is very unbalanced, the HCL
\texttt{rainbow\_hcl()} (or also \texttt{qualitative\_hcl()}) is (by
design) balanced with respect to luminance.
\begin{CodeChunk}
\begin{CodeInput}
R> wheel <- function(col, radius = 1, ...)
+    pie(rep(1, length(col)), col = col, radius = radius, ...) 
R> par(mar = rep(0.5, 4), mfrow = c(2, 2))
R> wheel(rainbow(8))
R> wheel(rainbow_hcl(8))
R> wheel(desaturate(rainbow(8)))
R> wheel(desaturate(rainbow_hcl(8)))
\end{CodeInput}
\end{CodeChunk}

\begin{figure}[t!]
\centering
\includegraphics[width=0.5\textwidth]{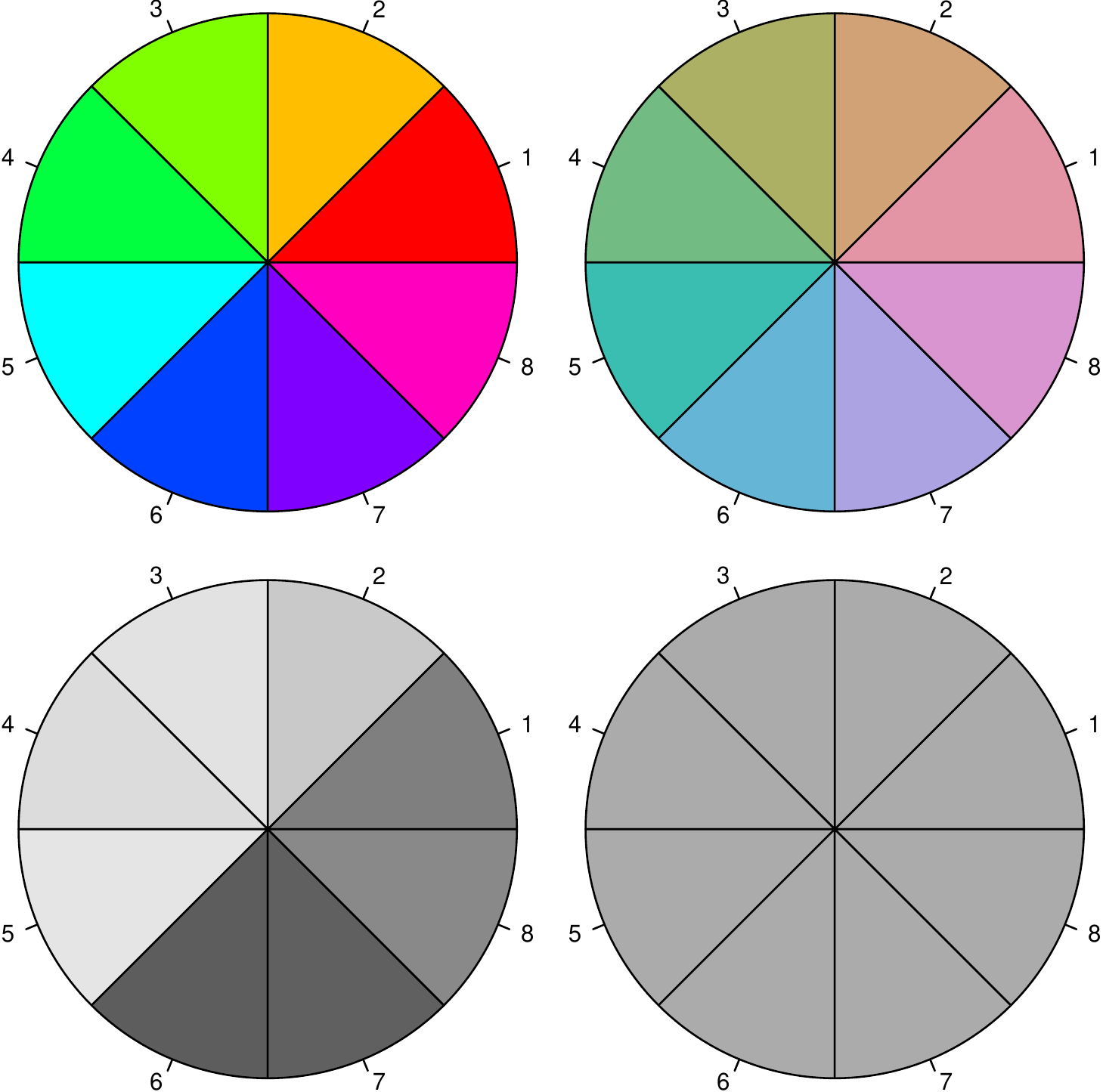} 
\caption[Color wheels in RGB (left) and HCL (right) space in color (top) and desaturated grayscale (bottom)]{Color wheels in RGB (left) and HCL (right) space in color (top) and desaturated grayscale (bottom).}\label{fig:desaturate-wheel}
\end{figure}

\subsection{Lighten or darken colors}\label{lighten-or-darken-colors}

In principle, a similar approach for lightening and darkening colors can
be employed as for desaturation above. The colors can simply be
transformed to HCL space and then the luminance can either be decreased
(turning the color darker) or increased (turning it lighter) while
preserving the hue and chroma coordinates. This strategy typically works
well for lightening colors, although in some situations the result can
be somewhat too colorful. Conversely, when darkening rather light colors
with little chroma, this can result in rather gray colors.

In these situations, an alternative might be to apply the analogous
strategy in HLS space which is frequently used in HTML style sheets.
However, this strategy may also yield colors that are either too gray or
too colorful. A compromise that sometimes works well is to adjust the
luminance coordinate in HCL space but to take the chroma coordinate
corresponding the HLS transformation.

We have found that typically the HCL-based transformation performs best
for lightening colors and this is hence the default in
\texttt{lighten()}. For darkening colors, the combined strategy often
works best and is hence the default in \texttt{darken()}. In either case
it is recommended to try the other available strategies in case the
default yields unexpected results.

Regardless of the chosen color space, the adjustment of the \texttt{L}
component by a certain \texttt{amount} can occur by two methods,
relative (the default) or absolute. For example for darkening these
either use \texttt{L\ -\ 100\ *\ amount} (absolute) or
\texttt{L\ *\ (1\ -\ amount)} (relative). See \texttt{?lighten} and
\texttt{?darken} for more details.

For illustration the qualitative palette suggested by
\citet{color:Okabe+Ito:2008} is transformed by two levels of both
lightening and darkening, respectively.
\begin{CodeChunk}
\begin{CodeInput}
R> oi <- c("#61A9D9", "#ADD668", "#E6D152", "#CE6BAF", "#797CBA")
R> swatchplot(
+    "-40%" = lighten(oi, 0.4),
+    "-20%" = lighten(oi, 0.2),
+    "  0%" = oi,
+    " 20%" =  darken(oi, 0.2),
+    " 40%" =  darken(oi, 0.4),
+    off = c(0, 0)
+  )
\end{CodeInput}
\end{CodeChunk}

\begin{figure}[t!]
\centering
\includegraphics[width=0.5\textwidth]{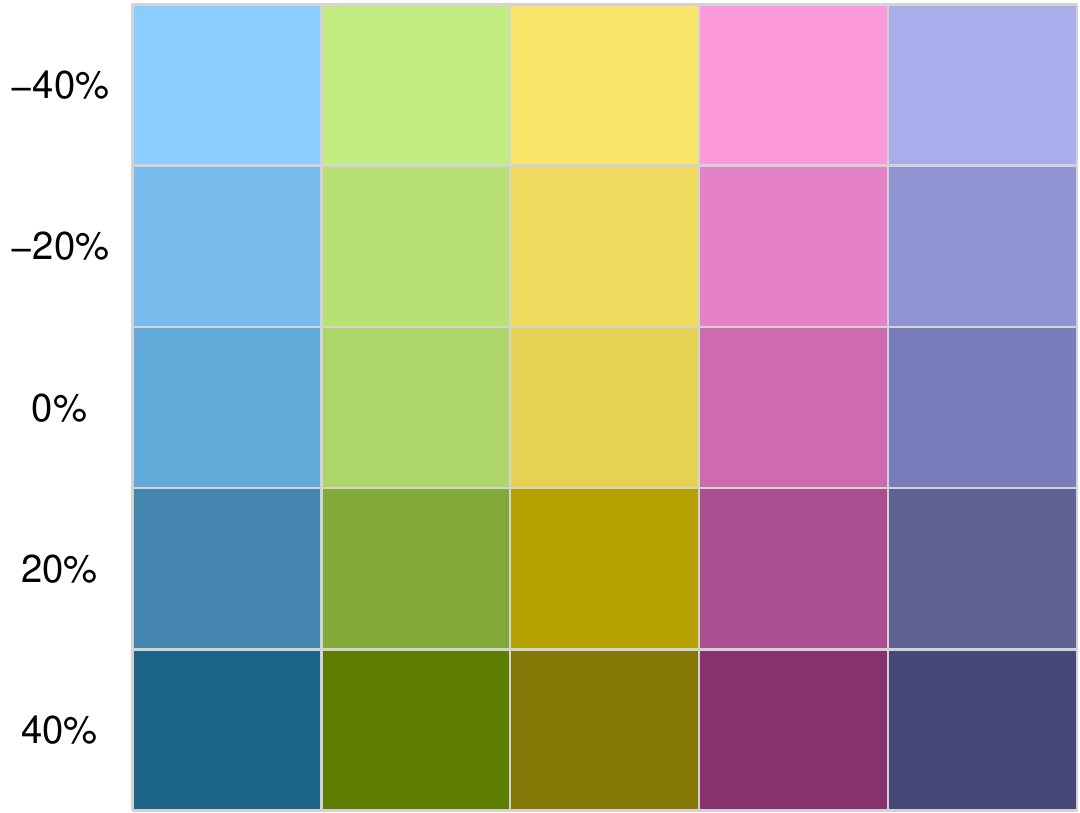} 
\caption[Okabe-Ito palette (0\%) along with two levels of both lightening and darkening, respectively]{Okabe-Ito palette (0\%) along with two levels of both lightening and darkening, respectively.}\label{fig:lighten-darken}
\end{figure}

\subsection{Maximum chroma for given hue and luminance}\label{maximum-chroma-for-given-hue-and-luminance}

As the possible combinations of chroma and luminance in HCL space depend
on hue, it is not obvious which trajectories through HCL space are
possible prior to trying a specific HCL coordinate by calling
\texttt{polarLUV()}. To avoid having to fix up the color upon conversion
to RGB \texttt{hex()} color codes, the \texttt{max\_chroma()} function
computes (approximately) the maximum chroma possible. For illustration
we show that for given luminance (here: \(L = 50\)) the maximum chroma
varies substantially with hue:
\begin{CodeChunk}
\begin{CodeInput}
R> max_chroma(seq(0, 360, by = 60), 50)
\end{CodeInput}
\begin{CodeOutput}
[1] 137.96  59.99  69.06  39.81  65.45 119.54 137.96
\end{CodeOutput}
\end{CodeChunk}
Similarly, maximum chroma also varies substantially across luminance
values for a given hue (here: \(H = 120\), green):
\begin{CodeChunk}
\begin{CodeInput}
R> max_chroma(120, seq(0, 100, by = 20))
\end{CodeInput}
\begin{CodeOutput}
[1]   0.00  28.04  55.35  82.79 110.28   0.00
\end{CodeOutput}
\end{CodeChunk}

\subsection{Additive mixing of two colors}\label{additive-mixing-of-two-colors}

In additive color models like \texttt{RGB()} or \texttt{XYZ()} it can be
useful to combine colors by additive mixing. Below a fully saturated red
and green are mixed, yielding a medium brownish yellow.
\begin{CodeChunk}
\begin{CodeInput}
R> R <- RGB(1, 0, 0)
R> G <- RGB(0, 1, 0)
R> Y <- mixcolor(0.5, R, G)
R> Y
\end{CodeInput}
\begin{CodeOutput}
       R   G B
[1,] 0.5 0.5 0
\end{CodeOutput}
\end{CodeChunk}

\section{Summary and discussion}\label{sec:summary}

This manuscript provides an overview of the broad capabilities of the
\pkg{colorspace} package for selecting individual colors or color
palettes, manipulating these colors, and employing them in various kinds
of visualizations.

In particular, the package provides various qualitative, sequential, and
diverging palettes derived by relatively simple trajectories in HCL
(Hue--Chroma--Luminance) space. In contrast to many other packages
providing modern balanced color palettes (such as \pkg{ColorBrewer.org},
\pkg{CARTO}, \pkg{viridis}, or \pkg{scico}) special emphasis is given to
flexibility of the palettes that can be adjusted to the particular needs
of a given data visulization. The manuscript also provides various tips
and tricks for choosing an effective palette in a given situation.
Further useful guidance is provided in many sources, including:
\citet{color:Ware:1988}, \citet{color:Okabe+Ito:2008},
\citet{color:Aigner:2010}, \citet{color:Stauffer+Mayr+Dabernig:2015},
\citet{color:Zhang:2015}, \citet{color:Rost:2018},
\citet{color:Wilke:2019}, and \citet{color:Ciechanowski:2019}, among
many others.

Further \proglang{R} packages that can complement the palettes provided
by \pkg{colorspace} include: \pkg{Polychrome}
\citep{color:Polychrome, color:Polychrome2} implements strategies for
qualitative palettes with many ``categories''. While the qualitative
palettes in Section~\ref{sec:hcl_palettes} yield only about 6--8 clearly
distinguishable colors due to the fixed chroma and luminance,
\pkg{Polychrome} relaxes this restriction and can thus find a larger
number of colors in CIELUV space that are spaced as far apart as
possible. It also has several functions for visualizing color palettes
in different ways. The palette collection packages \pkg{pals}
\citep{color:pals} and \pkg{paletteer} \citep{color:paletteer} also
provide a wide range of prespecified palettes, including some
qualitative schemes with many categories. Note that the palettes are
quite diverse, though, and not all of them are equally suitable for
coding qualitative information. The visualization functions in
\pkg{colorspace} from Section~\ref{sec:palette_visualization} may be
helpful in assessing their properties. \pkg{roloc}
\citep{color:roloc, color:roloc2} also provides color conversions, but
not between numeric colour spaces; instead, from numeric color spaces to
English color names.

In addition to the \proglang{R} version of \pkg{colorspace}, a
\proglang{Python}~2/\proglang{Python}~3 re-implementation is in beta and
available at \url{https://github.com/retostauffer/python-colorspace}. In
the manuscript we focus on the more mature \proglang{R} implementation
but replication materials for most examples are not only provided for
\proglang{R} but also for \proglang{Python}.

\section*{Computational details}

The results in this paper were obtained using \proglang{R} 3.5.2
\citep{color:R} with the packages \pkg{colorspace} 1.4.1
\citep{color:colorspace}, \pkg{ggplot2} 3.1.0 \citep{color:ggplot2pkg},
\pkg{RColorBrewer} 1.1.2 \citep{color:RColorBrewer}, \pkg{rcartocolor}
0.0.22 \citep{color:rcartocolor}, \pkg{viridis} 0.5.1
\citep{color:viridis}, \pkg{scico} 1.1.0 \citep{color:scico}.
\proglang{R} itself and all packages used are available from the
Comprehensive \proglang{R} Archive Network (CRAN) at
\url{https://CRAN.R-project.org/}.

\bibliography{colorspace}

\end{document}